\documentclass[fleqn, usenatbib]{mnras}

\usepackage{newtxtext,newtxmath}
\usepackage[T1]{fontenc}
\DeclareRobustCommand{\VAN}[3]{#2}
\let\VANthebibliography\thebibliography
\def\thebibliography{\DeclareRobustCommand{\VAN}[3]{##3}\VANthebibliography}

\usepackage{graphicx}
\usepackage{rotating}
\usepackage{multirow}
\usepackage{booktabs}
\usepackage{xcolor}

\usepackage{xspace}

\newcommand{\nustar}{{\it NuSTAR}\xspace}
\newcommand{\nicer}{{\it NICER}\xspace}

\newcommand{\xmm}{{\it XMM-Newton}\xspace}
\newcommand{\ergs}{erg s$^{-1}$\xspace }
\newcommand{\source}{SWIFT~J1749.4--2807\xspace}
\newcommand{\rg}{R$_{\rm G} $ }

\newcommand\T{\rule{0pt}{2.6ex}}       % Top strut
\newcommand\B{\rule[-1.2ex]{0pt}{0pt}}

\title[Spectral analysis of Swift J1749.4--2807]{Outflows and spectral evolution in the eclipsing AMXP SWIFT J1749.4--2807 with {\it NICER}, {\it XMM-Newton} and {\it NuSTAR}}

\author[A. Marino et al.]{
A. Marino$^{1,2,3,4}$\thanks{E-mail: marino@ice.csic.es}, A. Anitra$^3$, S. M. Mazzola$^{5}$, T. Di Salvo$^{3}$, A. Sanna$^{5}$, P. Bult$^{6,7}$, S. Guillot$^{8,9}$
\newauthor 
G. Mancuso$^{10,11}$, M. Ng$^{12}$, A. Riggio$^{5,4}$, A. C. Albayati$^{13}$, D. Altamirano$^{13}$, Z. Arzoumanian$^{7}$, L. Burderi$^{5}$, 
\newauthor 
C. Cabras$^{5}$, D. Chakrabarty$^{12}$, N. Deiosso$^{5}$, K. C. Gendreau$^{7}$, R. Iaria$^{3}$, A. Manca$^{5}$, T. E. Strohmayer$^{7,14}$
\\
\\
$^{1}$ Institute of Space Sciences (ICE, CSIC), Campus UAB, Carrer de Can Magrans s/n, E-08193 Barcelona, Spain \\
$^{2}$ Institut d'Estudis Espacials de Catalunya (IEEC), E-08034 Barcelona, Spain \\
$^{3}$ Universit\`a degli Studi di
  Palermo, Dipartimento di Fisica e Chimica, via Archirafi 36 - 90123 Palermo, Italy \\
$^{4}$ INAF/IASF Palermo, via Ugo La Malfa 153, I-90146 - Palermo, Italy \\
$^5$ Universit\`a degli Studi di Cagliari, Dipartimento di Fisica, SP Monserrato-Sestu km 0.7, I-09042 Monserrato, Italy \\
$^{6}$ Department of Astronomy, University of Maryland, College Park, MD 20742, USA \\
$^{7}$ Astrophysics Science Division, NASA Goddard Space Flight Center, Greenbelt, MD 20771, USA \\
$^{8}$ CNRS, IRAP, 9 avenue du Colonel Roche, BP 44346, F-31028 Toulouse Cedex 4, France \\
$^{9}$ Universit\'e de Toulouse, CNES, UPS-OMP, F-31028 Toulouse, France \\
$^{10}$Instituto Argentino de Radioastronom\'{\i}a (CCT-La Plata, CONICET; CICPBA), C.C. No. 5, 1894 Villa Elisa, Argentina \\
$^{11}$Facultad de Ciencias Astron\'omicas y Geof\'{\i}sicas, Universidad Nacional de La Plata, Paseo del Bosque s/n, 1900 La Plata, Argentina \\
$^{12}$ MIT Kavli Institute for Astrophysics and Space Research, Massachusetts Institute of Technology, Cambridge, MA 02139, USA \\
$^{13}$ School of Physics and Astronomy, University of Southampton, Southampton, SO17 1BJ, UK \\
$^{14}$ Joint Space-Science Institute, NASA’s Goddard Space Flight Center, Greenbelt, MD 20771, USA \\
}
\date{Accepted XXX. Received YYY; in original form ZZZ}

\pubyear{2022}

\begin{document}
\label{firstpage}
\pagerange{\pageref{firstpage}--\pageref{lastpage}}
\maketitle

\begin{abstract}
The neutron star low-mass X-ray binary \source is the only known eclipsing accreting millisecond X-ray pulsar. In this manuscript we perform a spectral characterization of the system throughout its 2021, two-week-long outburst, analyzing 11 \nicer observations and quasi-simultaneous \xmm and \nustar single observations at the outburst peak. The broadband spectrum is well-modeled with a black body component with a temperature of $\sim$0.6~keV, most likely consistent with a hot spot on the neutron star surface, and a Comptonisation spectrum with power-law index $\Gamma \sim 1.9$, arising from a hot corona at $\sim$12~keV. No direct emission from the disc was found, possibly due to it being too cool. A high truncation radius for the disc, i.e., at $\sim$20--30 \rg , was obtained from the analysis of the broadened profile of the Fe line in the reflection component. The significant detection of a blue-shifted Fe XXVI absorption line at $\sim$7~keV indicates weakly relativistic X-ray disc winds, which are typically absent in the hard state of X-ray binaries. By comparing the low flux observed during the outburst and the one expected in a conservative mass-transfer, we conclude that mass-transfer in the system is highly non-conservative, as also suggested by the wind detection. Finally, using the \nicer spectra alone, we followed the system while it was fading to quiescence. During the outburst decay, as the spectral shape hardened, the hot spot on the neutron star surface cooled down and shrank, a trend which could be consistent with the pure power-law spectrum observed during quiescence.  %Finally, we confirmed the claim that the mass-transfer in the system was highly non-conservative including the average bolometric flux calculated in this outburst.
%In all visits, the spectrum could be described with a black body component at $\sim$0.4-0.6~keV, most likely consistent with a hot spot on the Neutron Star surface, and a Comptonisation spectrum with power-law index $\Gamma \sim 1.8-2.1$. 
%This is a simple template for authors to write new MNRAS papers.
%The abstract should briefly describe the aims, methods, and main results of the paper.
%It should be a single paragraph not more than %250 words (200 words for Letters).
%No references should appear in the abstract.
\end{abstract}

\begin{keywords}
accretion, accretion discs -- stars:neutron -- X-rays: binaries -- X-rays, individuals: Swift J1749.4-2807
\end{keywords}

%%%%%%%%%%%%%%%%%%%%%%%%%%%%%%%%%%%%%%%%%%%%%%%%%%

%%%%%%%%%%%%%%%%% BODY OF PAPER %%%%%%%%%%%%%%%%%%

\section{Introduction}
Accreting millisecond X-ray pulsars, hereafter AMXPs, are low-mass X-ray binaries harbouring X-ray pulsars spinning at frequencies of about hundreds of Hz \citep[see, for reviews ][]{Patruno2021,Campana2018,DiSalvo2020}. Since the discovery of the first AMXP \citep[SAX J1808.4-3658,][]{Wijnands1998}, millisecond X-ray pulsations have been found in 24 other systems,  the most recent ones being MAXI J1816-195 \citep{Bult2022_atel} and MAXI J1957+032 \citep[][]{Ng2022_atel}. All AMXPs are transients and a large number of them have been seen only once in outburst and for relatively short periods. However, a few systems stand out for having been observed quite often in outburst \citep[e.g., SAX~J1808.4-3658, which in 2019 went in outburst for the eighth time,][]{Bult2020} or over time scales of years \citep[e.g., HETE~J1900.1-2455 or MAXI~J0911-655,][]{Patruno2012_hete,Sanna2017b}. Almost all AMXPs have short, i.e., $\sim$~hrs or mins, orbital periods and are therefore characterised by compact orbits, which can accommodate only very low mass companion stars, typically below 0.5 $M_{\odot}$. From a spectral point of view, the emission from these systems in outburst is usually dominated by the Comptonisation spectrum from a hot corona with electron temperatures $kT_{\rm e}$ usually of tens of keV \citep{DiSalvo2020}. The disc and the neutron star / boundary layer spectra are usually found with relatively low temperatures, i.e., below 1~keV. According to the canonical subdivision in hard and soft spectral states in X-ray binaries \citep[see, e.g., ][ for a review]{Done2007}, AMXPs in outburst are therefore (almost) always in hard state, with SAX J1748.9-2021 \citep{Pintore2016} and SAX J1808.4-3658 \citep[][]{DiSalvo2019} as perhaps the only sources observed to transition into the soft state. An additional spectral component arises from the Comptonisation spectrum emitted by the corona which hits the disc and it is thereby reprocessed by it; this component is dubbed reflection spectrum and it is a very common ingredient in X-ray binaries spectra \citep{Fabian1989}. \\ Reflection components are characterised by a forest of emission lines, the most remarkable one being the iron K line at 6.4--6.7~keV, and an excess at 10--30~keV called Compton hump. A proper investigation of these spectral features provides the opportunity to get insights on, e.g., the inner radius of the accretion disk, the ionization of the plasma in the disk, the inclination of the system. %Furthermore, even in outburst, these systems are quite faint, and attain X-ray luminosities usually between 10$^{36}$-10$^{37}$ erg s$^{-1}$. 
Reflection features have been found in most AMXPs \citep[but not all, see e.g., ][]{Miller2003,Falanga2005_1807,Sanna2018_16597,Sanna2018_1737}, for which data with high to moderate energy resolution were available \citep[see, e.g., ][]{Papitto2010_17511, Papitto2013_hete, Sanna2017}. Interestingly, the presence of disc winds, a common ingredient in X-ray spectra of high inclination X-ray Binaries (XRBs) during their soft state \citep[][]{Ponti2012}, has been established for one AMXP, i.e., IGR~J17591--2342 \citep[][]{Nowak2019}, while tentative detections have been obtained also for SAX J1808.4-3658 \citep{DiSalvo2019} and IGR~J17062--6143 \citep[e.g., ][]{VanDenEijnden2018}. Finally, most of these systems are bursters, as they have at least once displayed a type-I X-ray burst. \\
All of these characteristics make AMXPs remarkably similar to another class of X-ray binaries, the Very Faint X-ray Transients \citep[VFXTs, ][]{Muno2005, King2006, Wijnands2006_GC}. These sources are known to exhibit fainter outbursts with respect to the typical ranges observed in X-ray binaries, with peak luminosity at about 10$^{36}$~\ergs and even lower \citep[e.g., ][]{DelSanto2007, IntZand2009, Degenaar2017_VFXT}. The origin of such peculiar behaviour is not established yet \citep[for a discussion on the proposed explanation, see, e.g., the introduction of][]{Bahramian2021},  because of the relatively scarce  data from these objects. %Indeed, their low luminosity, usually below the detection threshold of All-Sky Monitors, such as the Burst Alert Telescope onboard the {\it Neil Gehrels Swift} Observatory \citep{gehrels04} or MAXI, combined with their transient nature makes catching these elusive objects in outburst very challenging. Many systems in the class are rather compact and have very degenerate or evolved companion stars, i.e., helium dwarfs or white dwarfs \citep{Intzand2007,Hameury2016}. Type-I X-ray bursts are also commonly observed and prove that a large majority of the systems in this class is formed by X-ray binaries hosting NSs. 
In a specific sub-class of the VFXTs family, the burst-only systems, the source is only detected during the type-I X-ray burst, as its persistent emission stays below the detection threshold of the All-Sky Monitors active at the time of the observation \citep{Cocchi2001,Cornelisse2002,Campana2009}. Even if AMXPs are usually not as faint as VFXTs, the latter ones sometimes display brighter outbursts, i.e., at 10$^{36}$--10$^{37}$ erg s$^{-1}$  \citep[the "hybrid" VFXTs, ][]{Delsanto2010,Marino2019b}. Furthermore, a number of systems have been already identified as both AMXPs and VFXTs. An example can be found among the so-called transitional millisecond pulsars or tMSPs \citep[see, for a recent review, ][]{Papitto2020}. This small group consists of three binary millisecond pulsars which switch between a rotation-powered state, where they appear as radio pulsars, and an accretion-powered state \citep{Archibald2009,Archibald2013,DiMartino2013,Bassa2014,Papitto2015,Papitto2019}. The typical X-ray luminosity shown by tMSPs in their "disc state", i.e., about 10$^{34}$~\ergs or lower, is comparable to the ones shown by VFXTs and it has been indeed proposed that all VFXTs may be tMSPs in active state \citep{Heinke2015}. Despite being usually found in such a sub-luminous state, in at least one tMSP a brighter, AMXP-like, outburst has been observed \citep[][]{Papitto2015}, somehow bridging the gap between tMSPs and AMXP. Finally, three systems who displayed VFXT activity in the past have been later identified also as AMXPs: IGR~J17494-3030 \citep{Ng2021}, MAXI J1957+032 \citep[][]{Ng2022_atel} and Swift~J1749.4--2807 (see below). 

\subsection{Swift J1749.4--2807}\label{ss:source}
\source was discovered in 2006 during a type-I X-ray burst \citep{Wijnands2009} and initially mistaken for a $\gamma$-ray burst \citep{Schady2006}. The following  X-ray activity of the source was very peculiar as the system reached the peak at a luminosity of about 5$\times$10$^{35}$\ergs (for a 7~kpc distance) and faded to quiescence within a day \citep{Wijnands2009, Campana2009}. The behaviour is consistent with the system being a burst-only source. By assuming that the observed burst reached the Eddington limit for a typical NS mass, an upper limit on the distance of $6.7\pm1.3$~kpc was posed. \source was observed again by the International Gamma-Ray Astrophysics Laboratory \citep[{\it INTEGRAL}, ][]{Winkler2003} in 2010, during its monitoring campaign of the Galactic Center \citep{Pavan2010, Ferrigno2011}. {\it Swift} and Rossi X-ray Timing Explorer \citep[{\it RXTE}, ][]{Bradt1993} Target of Opportunity (ToO) observations were promptly triggered and X-ray pulsations at 518 Hz were discovered \citep{Altamirano2011_1749}, allowing \source to be classified as an AMXP. The orbital ephemeris were determined by \citealt{Belloni2010} and \citealt{Strohmayer2010}, who found that the system has an orbital period of 8.82~hrs. The {\it RXTE} light curve also showed X-ray eclipses \citep{Markwardt2010}, making \source the first, and so far only, eclipsing AMXP. The inclination of the binary is therefore well constrained in the range 76$^\circ$--77$^\circ$ \citep{Altamirano2011_1749}. Contrarily to the other AMXPs, for which the mass of the donor is usually unknown, the precise determination of the inclination enabled also the companion star mass to be constrained, i.e., in the range $\sim$0.5--0.8~M$_{\odot}$ (for a NS mass comprised between 1.0 and 2.2~M$_{\odot}$). In quiescence, the cooling of the NS was followed in X-rays by \cite{Degenaar2012} and an attempt of finding its NIR/optical counterpart was performed by \cite{Davanzo2011}. On 2021 March 1st, after almost 11 years, JEM-X onboard {\it INTEGRAL} caught \source in a bright outburst again  \citep{Mereminskiy2021_atel}. Coherent pulsations were detected also in this case, this time with \nicer \citep{Bult2021_atel}.

In this manuscript, we present a detailed broadband X-ray spectral analysis of this peculiar object during its latest outburst. In the first part of this work, we analyze the 0.8--50~keV broadband spectrum of the system using \xmm, \nustar and \nicer data (Sec. \ref{ss:broadband}), while in the second one we analyze individually eleven \nicer observations covering the short-lived outburst, from rise to decay (Sec. \ref{ss:nicer_monitoring}). The main results obtained in this work are discussed in Section \ref{sec:disc}, while summary and conclusions are presented in Section \ref{sec:concl}.

\section{Observations \& Data reduction}\label{sec:obs}

The 2021 outburst of \source was the object of a number of pointed observations from different X-ray telescopes. In this paper we exploit the data collected by {\it XMM-Newton}, \nicer and {\it NuSTAR}. A summary of the observations is reported in Table \ref{tab:obs}. More details on the data reduction for each instruments are given in the following subsections.

\begin{table}
    \centering
    \begin{tabular}{l l l l l }
         \hline
         \hline
         & ObsID & \multicolumn{2}{c}{Start Time} & Exposure  \T\\
         & & (UTC) & (MJD) & ks  \B\\
         \hline
         & \multicolumn{4}{c}{ \xmm } \T \B \\
         \hline
         & 0872392001 & 2021-03-04 & 59277.01 & 57.9 \T \B\\
         \hline
         & \multicolumn{4}{c}{\nustar} \T \B \\
         \hline
%         Epoch & ObsID & \multicolumn{2}{c}{Start Time} & Exposure & Ref.  \\
%         & & (UTC) & (MJD) & ks & \\
%        \hline
          & 90701310002 & 2021-03-04 & 59277.30 & 49.4  \T \B\\
         \hline
         & \multicolumn{4}{c}{ \nicer } \T \B\\
         \hline
         N01 & 4658010101 & 2021-03-01 & 59274.62  & 9.3  \T\\
         N02 & 4658010102 & 2021-03-02 & 59275.00 & 18.2  \\
         N03 & 4658010103 & 2021-03-03 & 59276.03 & 9.8  \\
         N04 & 4658010104$^\dagger$ & 2021-03-04 & 59277.01 & 11.6  \\
         N05 & 4658010105 & 2021-03-05 & 59278.04  & 17.0  \\
         N06 & 4658010106 & 2021-03-08 & 59281.20 & 5.3  \\
         N07 & 4658010107 & 2021-03-09 & 59282.04 & 11.3  \\
         N08 & 4658010108 & 2021-03-10 & 59283.01 & 12.0  \\
         N09 & 4658010109 & 2021-03-11 & 59284.11 & 8.6  \\
         N10 & 4658010110 & 2021-03-12 & 59285.01 & 7.4  \\
         N11 & 4658010111 & 2021-03-13 & 59286.17 & 4.5  \B\\
        \hline
        \hline
    \end{tabular}
    \caption{List of the \xmm, \nustar and \nicer observations of the source used in this work. With $^\dagger$ we indicated the \nicer observation quasi-simultaneous, i.e. taken within the same day, to \xmm and \nustar used in the broadband spectral analysis (Sec. \ref{ss:broadband}).}
    \label{tab:obs}
\end{table}
\subsection{XMM-Newton}

The \xmm observation was performed on March 4th, 2021 between 01:10:36 UTC and 02:26:17 UTC for a duration time of 57.9~ks.
The two MOS detectors of the European Photon Imaging Camera \citep[EMOS,][]{Turner_01} were operating in Imaging and Timing mode respectively, while the PN-type CCD detector \citep[EPN,][]{struder_01} was in Timing mode. The Reflecting Grating Spectrometer \citep[RGS, two modules,][]{herder_01} was functioning in standard spectroscopy mode.
We reprocessed the data using the \xmm Science Analysis Software (SAS) v18.0.0. We produced calibrated photon event files using reprocessing tools {\tt emproc}, {\tt epproc} and {\tt rgsproc} for EMOS, EPN and RGS data, respectively.
We verified the absence of flare events in the EPN data extracting the light curve in the 10--12~keV energy range and we checked out the presence of pile-up contamination for MOS1: we used the task {\tt epatplot} that, displaying the observed pattern distribution versus the expected one, allowed us to show how the pile-up fraction was not negligible. 
Therefore, we decided to use only EPN and MOS2 data for the analysis, as the count-rates of the collected data (40 counts/s and 19 counts/s) were below the pile-up threshold reported for the detectors (800 counts/s and 100 counts/s for EPN and MOS operating in timing mode, respectively).
We extracted the EPN 0.2--15~keV light curve considering  PATTERN$\leq$4, to extract only single and double good events, and FLAG=0 from a rectangular region which included the brightest columns of the detector (between RAWX$\geq$30 and RAWX$\leq$45). For the background we extracted the events from a region far away from the source, between RAWX$\geq$5 and RAWX$\leq$20. Meanwhile, for MOS2 data, we selected the columns in the interval 289$\leq$RAWX$\leq$327 for the source and we extracted background from a region of the outer CCDs which were functioning in imaging mode, as suggested by SAS Data Analysis Threads\footnote{\url{https://www.cosmos.esa.int/web/xmm-newton/sas-thread-mos-spectrum-timing}}.
To verify the presence of dips or eclipses, we extracted the 0.2--3~keV and the 3--10~keV EPN light curve and calculated the hardness ratio. We observed two complete eclipses between 1.05~ks and 3.42~ks, and between 32.80~ks and 35.10~ks from the start time, and a partial eclipse at the end of the observation (starting at $46.80$~ks).  With the same aim, we accumulated the MOS2 light curve with a bin time of 100 seconds considering PATTERN$\leq$0 and FLAG=0 observing one eclipse between 29.96~ks and 32.22~ks from the start time.  
Furthermore, 6 Type-I X-ray bursts were detected by both cameras during the observation.

Finally we used {\tt tabgtigen} task to create the good time interval (GTI) file for both bursts and eclipses, and the {\tt xmmselect} task to extract the 0.2--15~keV light curve and spectrum of the persistent emission, excluding the time intervals at which the bursts and the eclipses occurred, which will be investigated elsewhere (Mancuso et al., {\it in preparation}). Also, the RGS light curve of this observation (produced combining RGS1 and RGS2 light curves with {\tt rgslccorr}) showed bursts and eclipses, as observed in EPN data. Then, we used the same procedure to filter them and obtained the persistent spectrum running the task {\tt rgsproc} until the ``{\tt fluxing}" final stage. Then we combined the first order spectra of RGS1 and RGS2 through {\tt rgscombine} to obtain the total RGS spectrum and the related response matrix (rmf) and ancillary file (arf).

\subsection{NuSTAR}
\nustar observed the system on March 4th, 2021, for a total exposure of 49.4~ks. We reduced the data using the standard \texttt{Nustardas} task, incorporated in \textsc{Heasoft} (v. 6.26.1) and using the latest \textsc{CALDB} version available. The source was selected by means of a circular area of 100" radius, centered at the coordinates of the source, i.e. R.A. (J2000) = 17$^{\rm h}$49$^{\rm m}$31.940$^{\rm s}$, Dec (J2000) = $-$28$^{\circ}$08$^\prime$05.89$^{\prime\prime}$ \citep[][]{Roming2009}.  In order to take into account any background non-uniformity on the detector, we extracted the background spectra using four circles of $\sim$50" radii placed on different areas of the image characterised by having negligible contamination from the source. We then used \texttt{Nuproducts} to build spectra and light curves. Also in this case, eclipses and bursts (if present) were excluded when creating the final products. We used data from both the hard X-ray imaging telescopes on board {\it NuSTAR}, i.e., the focal plane mirrors (FPM) A and B. Finally, we did not sum the FPMA and FPMB spectra, but rather fitted them simultaneously by leaving a floating cross-normalization constant\footnote{In accordance with the guidelines from the \nustar team, see FAQ page, issue 19: \url{https://heasarc.gsfc.nasa.gov/docs/nustar/nustar_faq.html}}. \\
\subsection{NICER}
\nicer monitored the 2021 outburst of \source with almost daily cadence (see Table \ref{tab:obs}). In this work, we analysed the whole sample of these \nicer observations. Data were reduced using \texttt{nicerl2} task (\texttt{NICERDAS 2019-05-21 v006}): we set recommended calibration processes, standard screening and we added the \texttt{niprefilter2$\_$coltypes=base,3c50} parameter so that the \texttt{3C50} model can be used to derive background spectra later.
%, however, we verified that the products obtained with the last version of the software, were in line with the ones used in our analysis.}
%\simo{Questa frase sulla compatibilità va  capito come e dove inserirla correttamente secondo me, perchè siamo soggetti a critiche tipo "e perchè non avete usato i prodotti ricavati dall'ultima versione?". Quindi conviene circoscrivere il commento ad uno specifico aspetto e dire che abbiamo verificato che non cambia nulla (non ricordo dove stava il problema sinceramente, bkg? rmf?), ancora meglio se in una nota a piè pagina}
We extracted the cleaned event files, checking that all detectors were active during observations and excluding data from two of them (labelled 14 and 34), in order to reduce the detector noise. We accumulated light curves from all the observations using the {\tt xselect} tool, finding several eclipses and bursts that we excluded from our analysis. Then we selected the GTI using \texttt{NIMAKETIME} and applied them to the data via \texttt{NIEXTRACT-EVENTS}, selecting events with PI channel between 25 and 1200 (0.25–12.0~keV).We used the {\tt nibackgen3C50} tool to extract both the spectra of the source and the background from the cleaned event files, selecting the 2020 gain calibration.
%In order to fix the distortions due to the \nicer calibration uncertainties, we re-normalised the spectra by using the residuals of a power-law fit to the Crab Nebula \citep{Ludlam2018}. 
Finally, we exploited the public files \texttt{nixtiaveonaxis20170601v002.arf} and \texttt{nixtiref20170601v001.rmf} as Ancillary Response File and Redistribution Matrix File, respectively, retrievable from the {\it NICER} website\footnote{See \url{https://heasarc.gsfc.nasa.gov/docs/nicer/proposals/nicer_tools.html}.}. %As a background spectrum we used the public background file \texttt{nixtiback20190807.pi}, also available in the HEASARC archive. 

\section{Spectral Analysis}\label{sec:spectral}
\begin{figure}
\centering
\includegraphics[width=\columnwidth]{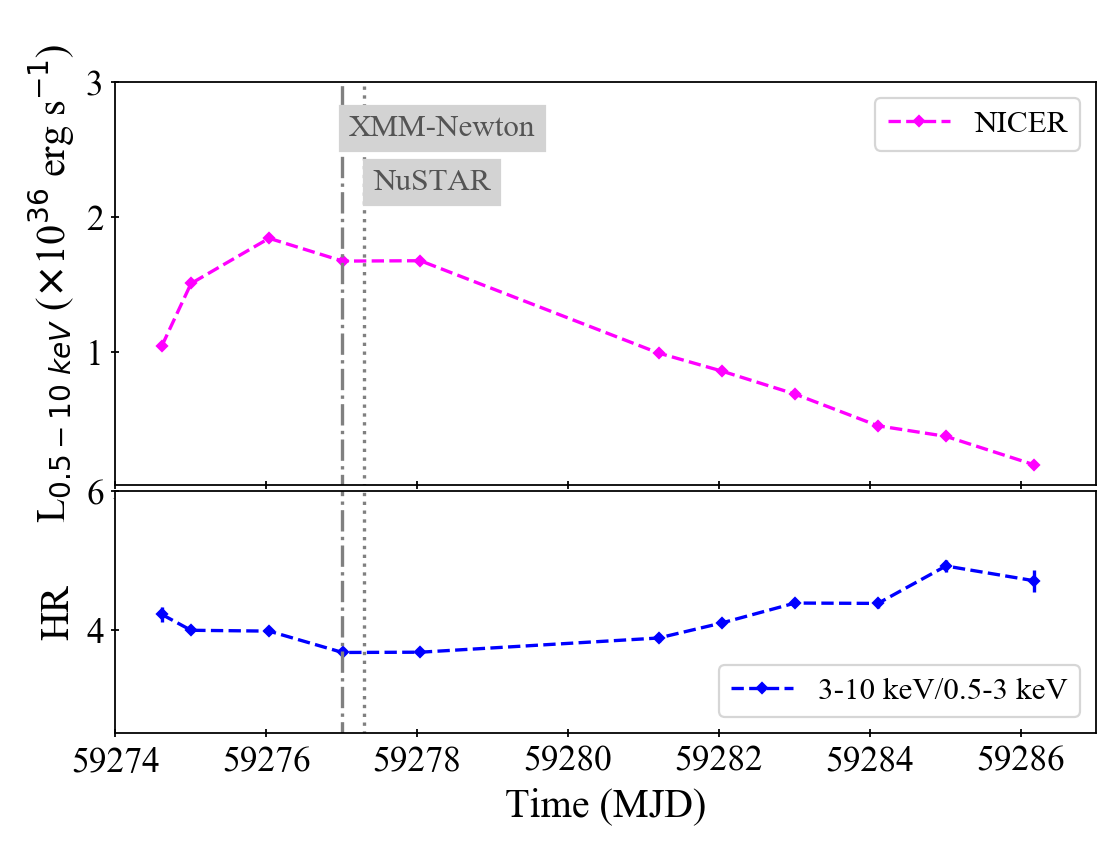}
\caption{({\it Top}) \nicer light curve of \source in X-ray (0.5-10~keV) luminosity (calculated for a 6.7~kpc distance) and relative Hardness Ratio (HR, {\it  bottom}). The times of the \xmm and \nustar observations are highlighted with dash-dotted and dotted gray lines respectively. The hardness values have been obtained by measuring the ratio between fluxes in the hard (3-10 keV) and soft (0.5-3 keV) bands, as estimated from the spectral analysis. }
\label{fig:nicer_lcurv}
\end{figure}

\begin{figure}
\centering
\includegraphics[width=\columnwidth]{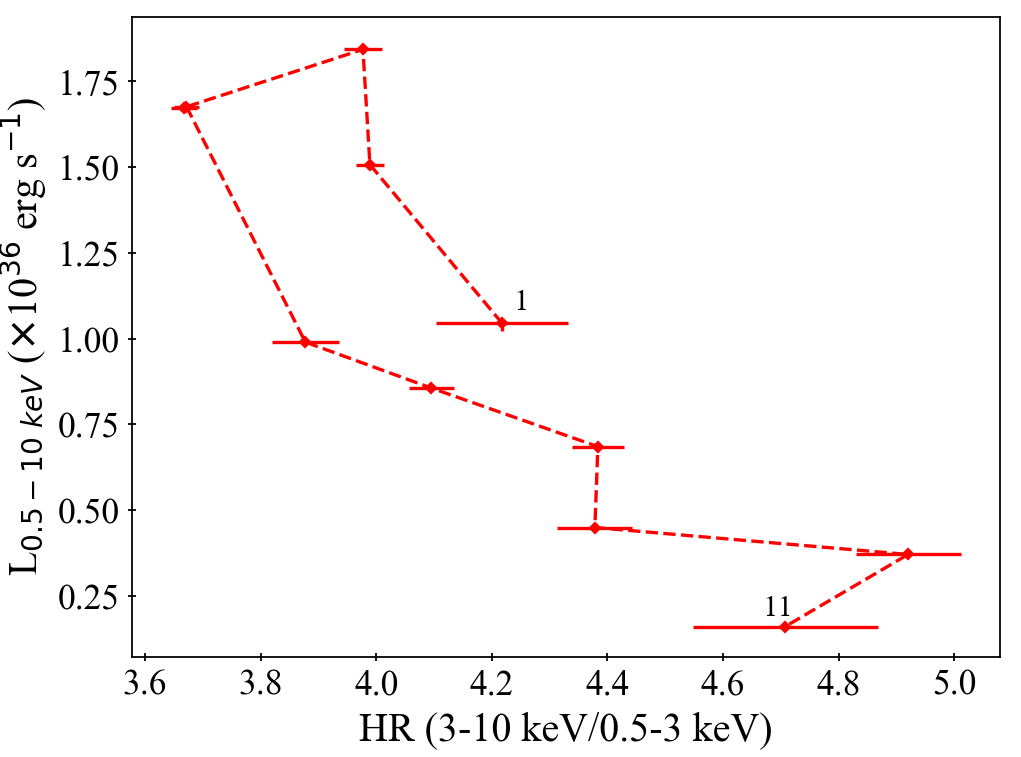}
\caption{\nicer Hardness Intensity Diagram (HID). The positions of the first and the last (eleventh) epochs are reported for clarity. The hardness values have been obtained by measuring the ratio between fluxes in the hard (3-10 keV) and soft (0.5-3 keV) bands, as estimated from the spectral analysis.}
\label{fig:nicer_hid}
\end{figure}

%The \nicer light curve of the system and the relative Hardness Intensity Diagram (HID) are shown in Fig. \ref{fig:nicer_lcurv} and \ref{fig:nicer_hid}, respectively. 
The \nicer light curve of the system is shown in Fig. \ref{fig:nicer_lcurv}. The \xmm and \nustar observations (highlighted in Fig. \ref{fig:nicer_lcurv} with vertical gray lines) have been taken the same day, around the peak of the outburst. In order to follow the spectral evolution of the system throughout the outburst, we analysed each of the \nicer spectra with \textsc{xspec} (see Sec. \ref{ss:nicer_monitoring} for more details on the spectral fitting procedure) and extracted by means of \texttt{cflux} the X-ray flux in three bands: 0.5--10 keV, 0.5--3 keV (soft band), 3--10 keV (hard band). We then plotted the 0.5--10 keV flux versus the hardness ratio to produce the Hardness Intensity Diagram (HID), shown here in Fig. \ref{fig:nicer_hid}. As apparent from the HID, the source remained quite spectrally stable throughout the whole outburst, with the hardness ranging only from 3.5 at the outburst peak to 5 at its culmination. \\
In the following subsections we proceed first (Sec. \ref{ss:broadband}) by analyzing a broadband spectrum including the \xmm and \nustar data, and the \nicer observation that was the closest in time with those pointings, i.e., ObsID 4658010104. In the \nustar spectrum we ignored data below 4~keV because of a mismatch between FPMA and FPMB, likely due to a known instrumental issue \citep{Madsen2020}. \nustar data higher than 50~keV were ignored as well,  as they appear background dominated. Similarly, RGS and \nicer data below 1 and 0.8~keV, respectively, were submerged by the instrument background and therefore neglected. EPIC-PN showed an excess below 1.5 keV, which was not present in \nicer and EPIC-MOS 2 data instead. Reports of similar excesses have been noticed several times in observations of bright objects performed by Epic-PN data in timing mode \citep[see, e.g., ][and references therein]{Dai2010,Egron2013} and sometimes ascribed to calibration issues. We therefore kept EPIC-PN data only between 2.4 and 10~keV, in order to ignore the 1.8~keV (Si K-edge) and 2.3~keV (Au M-edge) instrumental features as well. All the data used in this work were grouped exploiting the optimal binning recipe by \cite{Kaastra2016}, which allows to have a grouping reflecting the spectral resolution of the instrument in a given energy range and prevents any oversampling issue. \\
We then exploit the whole set of \nicer observations 
%to perform an analogous spectral analysis on a cumulative spectrum obtained by summing all of the 11 available \nicer data sets (Subsection \ref{ss:nicer_sum}). This strategy is justified by the apparent spectral stability of the system, as from Fig. \ref{fig:nicer_hid}. Notwithstanding its presumed lack of variability, we also 
to perform a round of analyses on each \nicer spectrum, hunting for possible evolution of the physical parameters of the system throughout the whole outburst. The results of such finer analysis are presented in Sec. \ref{ss:nicer_monitoring}. \\
We used {\sc XSPEC} v12.10.1f to perform the spectral fit. For each analysed observation, we included the \textsc{tbabs} component in the spectral model to take into account the effect of  the interstellar absorption, setting the photoelectric cross-sections and the element abundances to the values provided by \cite{Verner1996} and \cite{Wilms2000}, respectively. A \textsc{constant} component was also used to serve as cross-calibration constant.

\subsection{Broadband spectral analysis}\label{ss:broadband}
\subsubsection{Continuum analysis}
In order to probe the spectral shape of the source, we started by fitting the broadband \xmm - \nicer - \nustar spectrum with an absorbed \textsc{powerlaw} model. The fit residuals reveal  the presence of several structures, including a thermal component at low energies, i.e., below 2 keV, and the clear signature of a reflection Compton hump beyond $\sim$10 keV. We found that the thermal component could be satisfactorily described by a blackbody component from the NS/boundary layer (\textsc{bbodyrad}), a multi-colour disc blackbody \citep[\textsc{diskbb} or \textsc{diskpn}, e.g. ][]{Gierlinski1999} or a combination of the two models, but selected \textsc{bbodyrad} only as it gave the most physically reliable fit. Indeed, using only \textsc{bbodyrad}, we obtain a blackbody temperature $kT_{\rm bb}$ of $\sim$0.7 keV and a radius $R_{\rm bb}$\footnote{In \textsc{bbodyrad}, the normalisation parameter $K_{\rm bb}$ found by the spectral fit is connected to the blackbody radius $R_{\rm bb}$ by the formula $K_{\rm bb}=\left(R_{\rm bb}/D_{\rm 10 \ kpc}\right)^2$, with $D_{\rm 10 \ kpc}$ the distance of the source in units of 10~kpc.} of about $\sim$5 km, compatible with the presence of a hot spot on the NS surface. On the contrary, including only one disc component, as \textsc{diskbb} or alternatively a more sophisticated model such as \textsc{diskpn}, leads inevitably to unphysical results. A \textsc{diskbb} at a temperature of about 1~keV could in principle replace \textsc{bbodyrad}, but the relatively small normalization $K_{\rm disk}$ value obtained would translate\footnote{We converted the normalization into the inner disc radius by taking advantage of the relation: $K_{\rm diskbb}=\left( R_{\rm in}/D_{10}\right)^2\cos{i}$, with $K_{\rm diskbb}$ the \textsc{diskbb} normalization and $D_{10}$ the distance of the system in 10~kpc units, and using the colour correction factors $\kappa$=1.7 and a $\xi$ correction factor for the torque-free boundary
condition of 0.42 \citep{Kubota1998, Gierlinski2002}.} into a rather small inner radius of the disk, i.e., of $\sim$ 2~km, which is nonphysical for an accreting NS. Similarly, a single \textsc{diskpn} component, at a temperature of $\sim$1.0 keV and an inner radius of 50 \rg, provides an acceptable fit, but in order to explain the very low value found for the normalisation $K_{\rm diskpn}$ we need to invoke a distance $D$ of about 60 kpc\footnote{Since, $K_{\rm diskpn}$ can be expressed as $K_{\rm diskpn}=M_{\rm NS}\cos{i}/(D^2f^4)$, with $f$ the ratio between color and effective disc temperatures, we calculated the distance by assuming typical values for $M_{\rm NS} (1.5 M_\odot)$, $f$ (1.7), the known inclination of the system (77$^\circ$) and the value of $K_{\rm diskpn}$ obtained by the fit.}, almost 10 times higher than the distance range estimated for the system \citep[][]{Wijnands2009}. Keeping both \textsc{bbodyrad} and \textsc{diskbb} returns a statistically acceptable fit as well, but the normalization $K_{\rm disk}$ attained extremely high ($>2\times10^4$) values this time, which would require inner radii higher than 400~km, which is rather odd \citep[for the truncation radii typically found in this class of system, see e.g.][and references therein]{VanDenEijnden2017_Burster}. Using \textsc{diskpn} instead of \textsc{diskbb} does not solve the issue, as the distance $D$ required to justify the obtained normalisation value would be lower than 1 kpc. \\
In order to reproduce correctly both Comptonisation and reflection spectra, we replaced the \textsc{powerlaw} with \textsc{relxillCp} \citep{Garcia2014}. 
%\simo{Forse non elencherei tutti i parametri del modello, evidenzierei solo quelli fissati}
The main parameters of the model are therefore: the \textsc{bbodyrad} temperature $kT_{\rm bb}$ and normalization $K_{\rm bb}$, the disc emissivity $\epsilon$, the inclination of the system $i$, the inner radius of the disc $R_{\rm in}$, the electron temperature of the corona $kT_{\rm e}$, the ionization $\xi$, the iron abundance A$_{\rm Fe}$ and the reflection fraction $f_{\rm refl}$. The inclination of the system is known from the study of the eclipses, so that it was fixed to the value 76.9$^\circ$ \citep{Altamirano2011_1749}. In all the performed fits, we also froze $\epsilon$ to 3 \citep[a value commonly found in X-ray binaries, see e.g., ][]{Dauser2013,Marino2019b} and A$_{\rm Fe}$ to 1.0, since the fit was unable to find constraints on them. Finally, in order to account for small differences in calibration, we did not tie the power-law photon index $\Gamma$ between the \nicer and EPN and MOS 2 spectra. We checked systematically that discrepancies between the best fit values obtained for $\Gamma$ were always lower than 10\%. We notice that the seed photon temperature of the Comptonisation can not be changed and it is set to be very low, i.e. $kT_{\rm seed} \lesssim 0.1 \ {\rm keV}$, i.e. compatible with a cold, possibly truncated, disc. \\
The overall shape of the spectrum was well fitted with this model: $\chi^2_{\nu}$(d.o.f.)=1.38(670). According to this analysis, the photon index $\Gamma$ of the spectrum was about 1.9, the NS surface was emitting at a blackbody temperature of $\sim$ 0.7~keV, the blackbody radius was found to be $\sim$ 5~km, the corona had an electron temperature of 12--14~keV, the disc was truncated at a radius of about 20--34 \rg and the fraction of reflected photons was around 0.1. %These parameters are consistent with the source being in a spectral hard state, a common behaviour for AMXPs. 
%We tested whether the soft thermal component could be also described with a disc blackbody model (\textsc{diskbb} in \textsc{XSPEC}) or a combination of it with a simple blackbody model (\textsc{diskbb}+\textsc{bbodyrad}). While both alternatives could give acceptable fits, we still preferred using \textsc{bbodyrad} only since it returned the most physically motivated results, as discussed in Sec. \ref{ss:geometry
\subsubsection{Analysis of the absorption feature}
\begin{figure}
\centering
\includegraphics[width=\columnwidth]{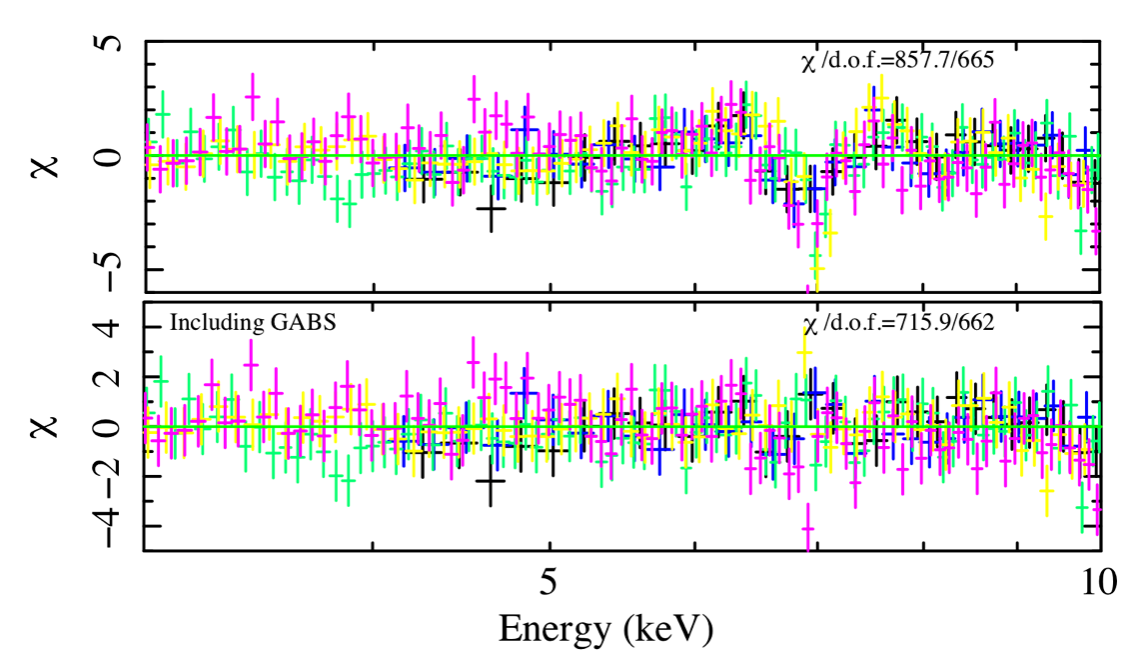}
\caption{Comparison between the residuals obtained with different models in the range 3--10 keV. Models tested: \texttt{tbabs}$\times$(\texttt{bbodyrad}+\texttt{relxillCp}), {\it top panel}; \texttt{tbabs}$\times$\texttt{gabs}$\times$(\texttt{bbodyrad}+\texttt{relxillCp}), {\it bottom panel}. Data: \nicer (green), {\it XMM-Newton}/EPIC-pn (yellow), EPIC-MOS2 (magenta), and {\it NuSTAR} (blue--black).}
\label{fig:residuals_focus}
\end{figure}

\begin{figure*}
\centering
\includegraphics[width=0.9\textwidth]{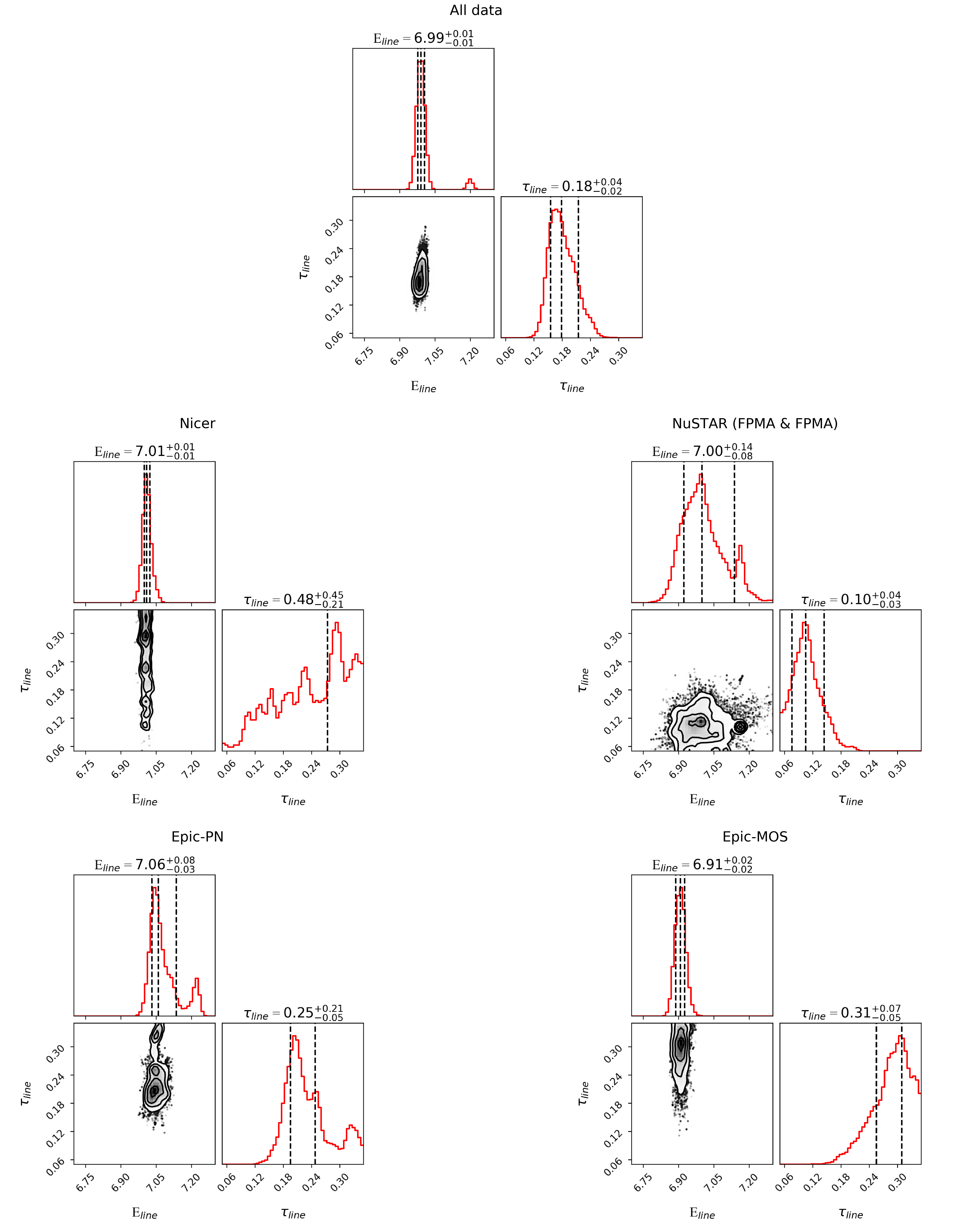}
\caption{Posterior probability distributions for $E_{\rm line}$ and $\tau_{\rm line}$ for the entire dataset (top panel), \nicer (middle left), \nustar (middle right), Epic-PN (bottom left) and Epic-MOS 2 (bottom right). Contours represent the 1$\sigma$ , 2$\sigma$ and 3$\sigma$ confidence levels. Marginal posterior distributions are shown as histograms with the median and 1 $\sigma$ intervals of confidence highlighted as dashed lines.}
\label{fig:contour_plots}
\end{figure*}

\begin{figure}
\centering
\includegraphics[width=\columnwidth]{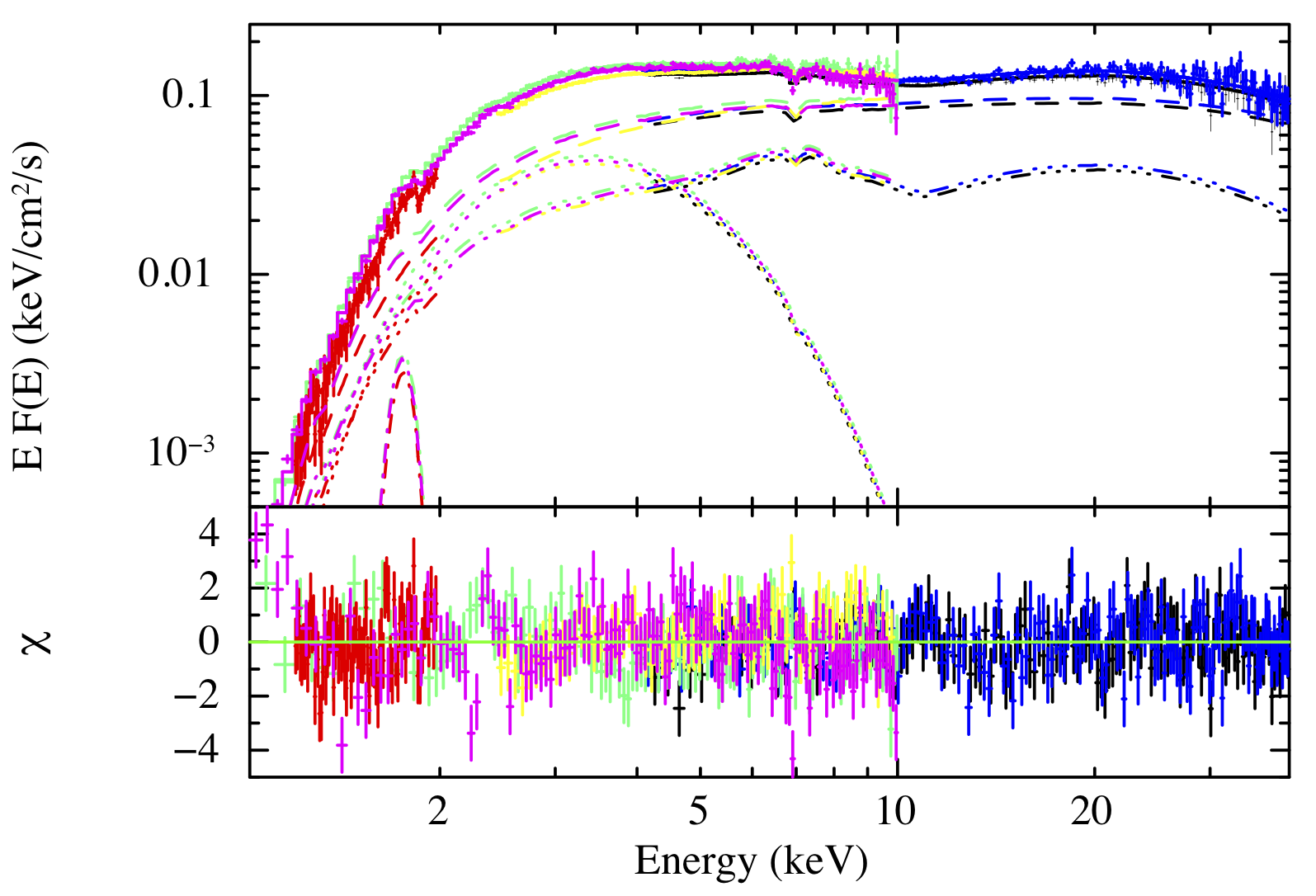}
\caption{Energy spectra, best-fit model and residuals for Epoch 4. Data: \nicer (green), {\it XMM-Newton}/EPIC-pn (yellow), EPIC-MOS2 (magenta), {\it XMM-Newton}/RGS (red) and {\it NuSTAR} (blue--black). Different line styles were adopted to distinguish between the different components: dot for \textsc{bbodyrad}, dash for \textsc{relxillCp} and,
dash-dot-dot-dot for \textsc{relxillCp} with the parameter $f_{\rm refl}$ set negative in order to highlight the pure reflection component, and dash-dot for the instrumental \textsc{gaussian} at 1.7~keV, due to Si fluorescence.}
\label{fig:residuals}
\end{figure}

Despite the acceptable fit, some local residuals still existed, as shown in Fig. \ref{fig:residuals_focus}. In particular, the presence of a pronounced dip around 7~keV clearly pointed out the presence of an absorption feature. In order to fit it, we initially included the multiplicative \textsc{gabs} component. The obtained value of $E_{\rm line}$ indicates that the feature is a slightly blueshifted absorption line from Fe XXVI, as commonly found for high inclination LMXBs \citep[see, e.g., ][]{Ponti2014}. As the detection of such a feature indicates the presence of ionised absorbing plasma in the system, we also included \textsc{zxipcf}, which takes into account the case where a fraction $f_{\rm abs}$ of the source is covered by absorbing photoionised matter, with ionization parameter $\xi_{\rm abs}$. The model was initially developed for AGNs \citep{Miller2007,Reeves2008}, but it has been applied to high inclination X-ray binaries as well \citep[see, e.g.,][]{Ponti2015,Iaria2020}. Besides of $f_{\rm IA}$ and $\xi_{\rm IA}$, the main parameter of the model is the equivalent hydrogen column of the material, $N_{\rm H, IA}$. The \textsc{zxipcf} component is in principle able to reproduce also the Fe XXVI line. However, in our case, keeping \textsc{gabs} in the model was necessary to fully clear the residuals \citep[see, e.g. ][ for a discussion]{Ponti2015}. We also attempted to replace \textsc{gabs} with a second \textsc{zxipcf} component but, in spite of a clear improvement in the residuals and in the fit, the parameters in the second \textsc{zxipcf} were left completely unconstrained by the fit. We therefore decided to use only one \textsc{zxipcf} component and \textsc{gabs} for the spectral analysis. \\
According to our results, the absorbing material decribed by \textsc{zxipcf} covers a fraction of 10\% or lower of the X-ray main source and it is characterised by a high ionisation ($\log{\xi_{\rm IA}}$ between 3 and 4.5) and high $N_{\rm H, IA}$ (lower limit of $\sim$3$\times$10$^{24}$ cm$^{-2}$). The Fe XXVI line was successfully fitted with \textsc{gabs} at an energy $E_{\rm line}$ of 6.99$^{+0.03}_{-0.02}$~keV, an associated optical depth at the center of about $\tau_{\rm line}\sim0.2$, and an equivalent width of$\sim$40 eV. Such a feature is found to be significantly blueshifted with respect to its rest-frame energy, i.e., 6.9662~keV \citep{Verner1996}, indicating an outflowing disc wind at a velocity $v_{\rm out}$ of $\sim$600--2700~km s$^{-1}$. In order to further investigate the line profile, we applied the Goodman-Weare algorithm of Monte Carlo Markov Chain \citep[MCMC;][]{Goodman2010} to produce contour plots for $E_{\rm line}$ and $\tau_{\rm line}$. We used 20 walkers and a chain length of 5$\times$10$^5$, to calculate the marginal posterior distributions of the best-fit parameters. The same procedure was applied to the spectra of \nicer , \nustar (including both FPMA and FPMB), Epic-PN and Epic-MOS 2 taken singularly, with the aim to explore how the feature is observed by the different instrument. As apparent, some level of discrepancy between the line profile seen by the different instruments has to be taken into account, especially with regards to Epic--MOS 2. Such a discrepancy could be a symptom of a residual level of pile-up or some other systematics in MOS 2. The results are presented in Figure \ref{fig:contour_plots}, where the MCMC chains are visualised by means of \texttt{corner.py} \citep[][]{Foreman2016}.
%Furthermore, the component was statistically highly significant, with probability of improvement by chance of 5$\times$10$^{-50}$ as estimated with \textsc{ftest}.
%By describing such feature with a \textsc{gaussian} component with negative normalization, 
%In order to fit it, we added a \textsc{gaussian} component, with negative normalization. The line was successfully fitted with this component, at an energy $E_{\rm line}$ of 7.00$\pm$0.02~keV, with equivalent width of$\sim$40 eV, detected at a 6$\sigma$ confidence.  (Sugg: forse meglio aggiungere  come è stata calcolata la confidence, è una domanda che a me i referee fanno sempre) \\
Finally, a second feature could be appreciated in emission around 1.7~keV in the \nicer residuals. Then, we added a second \textsc{gaussian} line (this time with positive normalization) and found it significant (5.5$\sigma$) at an E$_{\rm line,2}$ of 1.71~keV and a $\sigma_{\rm line,2}$ of $\sim$0.01~keV. This narrow feature is most likely a Si fluorescence line from the Focal Plane Modules (M. Corcoran, private communication). The final fit is shown in Fig. \ref{fig:residuals}, while the best-fit parameters are listed in detail in Table \ref{tab:fit_broadband}.

\begin{table}
\centering
%\caption{XRT+NuSTAR+BAT, two-reflection components model, m$_{\rm S}$ free}
\begin{tabular}{c c c c }
\hline 
%\hline
%\multicolumn{5}{l}{\textbf{Spectral analysis}} \\
\hline
 \multicolumn{4}{c}{\large \bf Broadband spectral analysis} \T \\
 \multicolumn{4}{c}{\nicer + \xmm + \nustar}  \B \\
 \hline
 \multicolumn{4}{c}{{\bf model:} \texttt{tbabs}$\times$\texttt{zxipcf}$\times$\texttt{gabs}$\times$(\texttt{relxillCp}+\texttt{bbodyrad})} \\
\hline
{\bf Model} & {\bf Parameter} \T \B \\
\cmidrule(lr){2-4} 
\textsc{tbabs} & $N_{\rm H}$ & $\left( 10^{22} \  {\rm cm}^{-2} \right)$ & {4.62$\pm$0.05}  \\
\cmidrule(lr){2-4} 
\\
%\cmidrule(lr){2-5} \cmidrule(lr){6-7} 
%\multirow{2}{*}{\textsc{diskbb}} & $kT_{\rm disk}$ & $\left( {\rm keV} \right)$ & $0.0452^{+0.008}_{-0.007}$ & 0.035$\pm$0.002 & &  $0.043\pm0.002$ \T\\
%& $K_{\rm disk}$ & $\left(\times 10^{13}\right)$ &  $>1$ & $>100$ & & $>1$ \B \\
%\cmidrule(lr){2-5} \cmidrule(lr){6-7}
%\\
\cmidrule(lr){2-4} 
\multirow{3}{*}{\textsc{zxipcf}} & $N_{\rm H, abs}$ & $\left( 10^{22} {\rm cm}^{-2} \right)$ & >280   \\
& $\log{\xi_{\rm abs}}$ & & 4.0$^{+0.6}_{-1.1}$\\
& f$_{\rm cov}$ & & 0.07$\pm$0.04\\
\cmidrule(lr){2-4} 
\\
\cmidrule(lr){2-4} 
\multirow{2}{*}{\textsc{bbodyrad}} & $kT_{\rm body}$ & $\left( {\rm~keV} \right)$ &  $0.695^{+0.013}_{-0.014}$  \\
& $R_{\rm bb}$ & $\left( {\rm km}\right)$ & 5.0$^{+1.5}_{-1.5}$  \\
\cmidrule(lr){2-4} 
\\
\cmidrule(lr){2-4} 
\multirow{10}{*}{\textsc{relxillCp}} & $\epsilon$ & & {(3.0)} \\
& i & $\left( ^\circ \right)$ & {(76.9)}  \\
& $R_{\rm in}$ & \rg &  {26$^{+8}_{-6}$}  \\
& $\Gamma$ & & 1.91$\pm$0.02  \\
& $kT_{\rm e}$ & $\left( {\rm keV} \right)$ & {12.9$^{+1.6}_{-1.1}$}   \\
& $\log{\left(\xi\right)}$ & & 3.30$^{+0.15}_{-0.13}$   \\
& A$_{\rm Fe}$ & & (1.0)  \\
& $f_{\rm refl}$ & & {0.16$^{+0.10}_{-0.04}$}  \\
& $K_{\rm RELX}$ & $\left(\times 10^{-4}\right)$ & {9.7$^{+1.3}_{-2.2}$}  \\
%\cmidrule(lr){2-5} \cmidrule(lr){6-7}
%\\
%\cmidrule(lr){2-5} \cmidrule(lr){6-7}
%\multirow{4}{*}{\textsc{NthComp}} & $\Gamma$ & & - & - & & $1.98^{+0.03}_{-0.02}$ \\
%& $kT_{\rm seed}$ & $\left( {\rm keV} \right)$ & - & - & & =$kT_{\rm disk}$ \\
%& $kT_{\rm e}$ & $\left( {\rm keV} \right)$ & - & - & & (15) \\
%& $K_{\rm NTC}$ & & - & - & & 0.092$\pm$0.005 \\
\cmidrule(lr){2-4} 
\\
\cmidrule(lr){2-4} 
\multirow{3}{*}{\textsc{gabs}} & $E_{\rm line}$ & (keV) & 6.99$^{+0.03}_{-0.02}$ \\
& $\sigma_{\rm line}$ & (keV) & 0.11$\pm$0.06\\
& $\tau_{\rm line}$ &  & 0.17$\pm$0.02 \\
\cmidrule(lr){2-4} 
\\
\hline
& $\chi^2_{\nu}$ & (d.o.f.) & {\bf 1.07(660)} \\
\hline
\hline
\end{tabular}
\caption{Fit results. Quoted errors reflect 90\% confidence level. The parameters that were kept frozen during the fits are reported between round parentheses.}
\label{tab:fit_broadband}
\end{table}
\subsection{\nicer timing analysis}\label{ss:nicer_analysis}
\begin{figure}
\centering
\includegraphics[width=0.9\columnwidth]{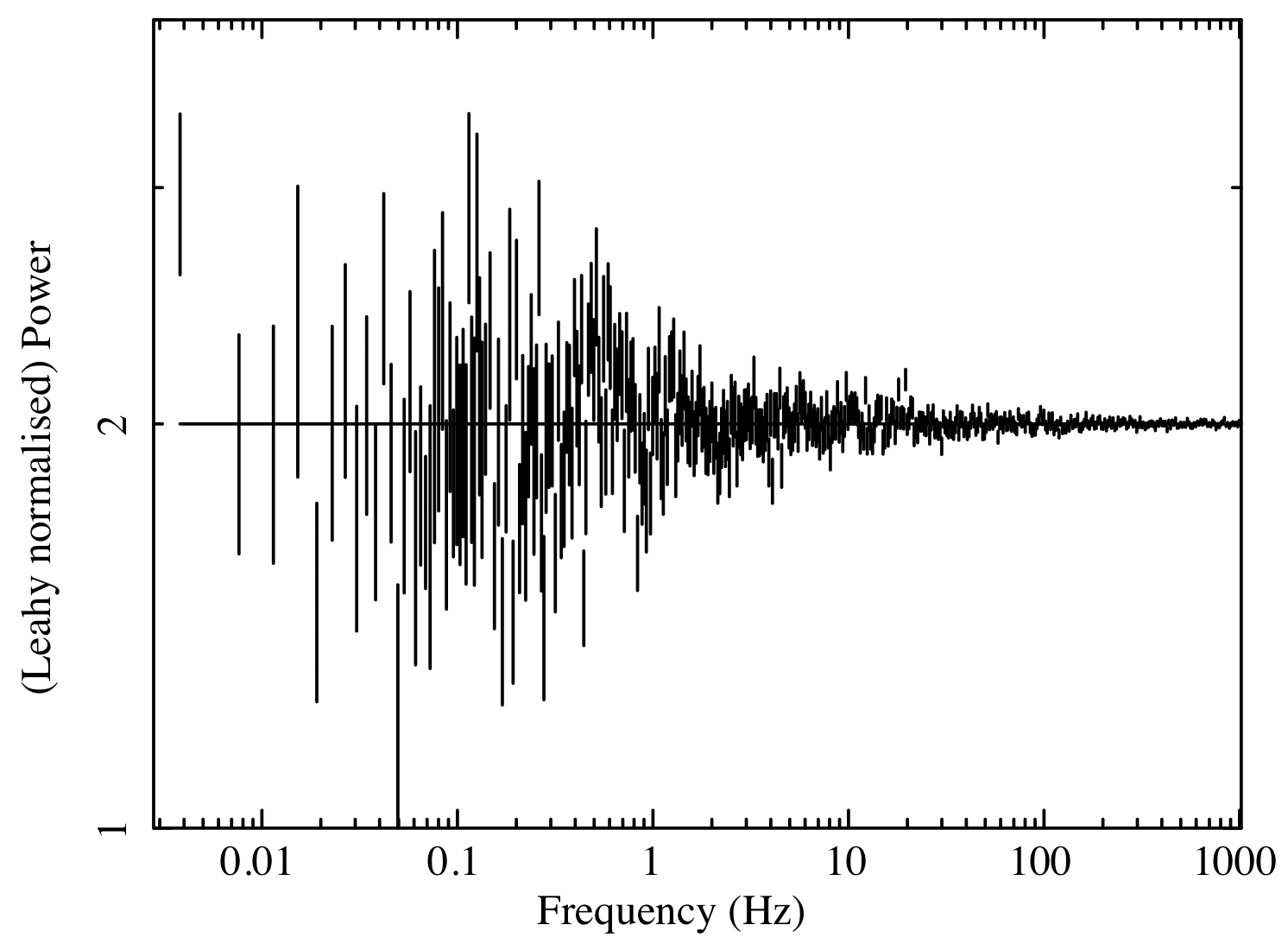}
\caption{\nicer power density spectrum for Epoch 4. A \textsc{cons} model at value 2.0 is displayed as an horizontal black line.}
\label{fig:power_spectrum}
\end{figure}

In order to corroborate the spectral state identification, we extracted the Leahy-normalised \nicer Fourier Power Density Spectrum (PDS) for Epoch 4 in the 0.5--10 keV energy range. We used a bin-time of 5 ms. We did not subtract the Poisson noise contribution, but rather fitted it with a constant component. The obtained PDS and its best-fit model are shown in Figure \ref{fig:power_spectrum}. The PDS is consistent with a constant of $\sim$2 at higher frequencies, compatible with Poisson noise. We obtained a value of about 6\% for the overall root-mean-square {\it rms}. The system shows therefore little to no X-ray variability, in contrast to what typically expected for X-ray binaries in a hard or hard-intermediate state. This will be further discussed in Section \ref{sec:disc}.

%power law-like continuum ($\propto \nu^{-\alpha}$, called very low frequency noise, VLFN) is usually observed below $\sim 1$ Hz in the softest states; for atoll sources in the end part of the BS (upper banana), and for Z sources on the FB. It has been variously ascribed to accretion-rate variations and unsteady nuclear burning. 

\subsection{\nicer monitoring}\label{ss:nicer_monitoring}
\begin{figure*}
\centering
\includegraphics[width=\textwidth]{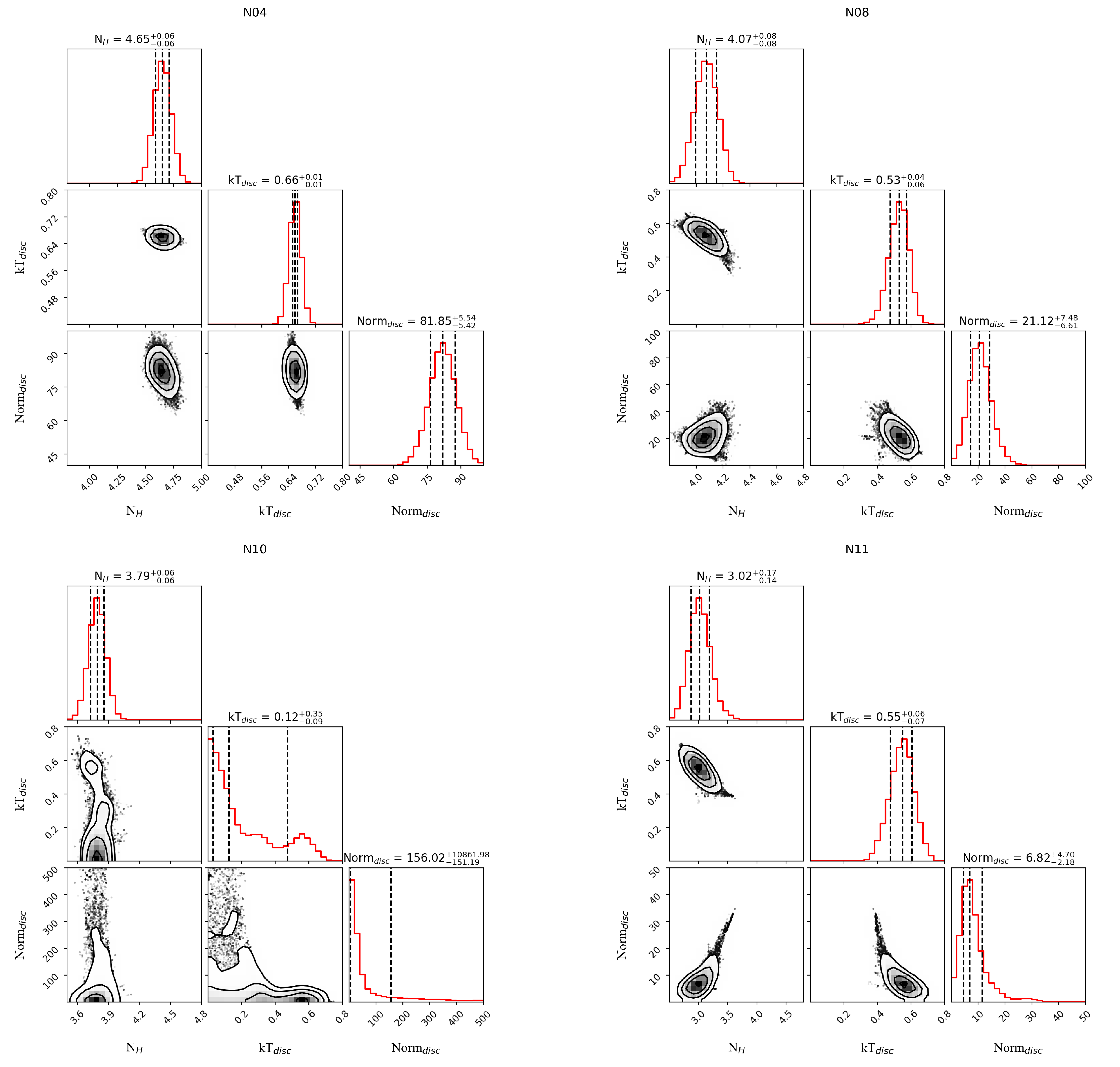}
\caption{Posterior probability distributions for $N_H$, $kT_{\rm bb}$ and the \texttt{bbodyrad} normalization ${\rm Norm}_{\rm bb}$. Contours represent the 1$\sigma$ , 2$\sigma$ and 3$\sigma$ confidence levels. Marginal posterior distributions are shown as histograms with the median and 1 $\sigma$ intervals of confidence highlighted as dashed lines.}
\label{fig:corner}
\end{figure*}

\begin{figure}
\centering
\includegraphics[width=\columnwidth]{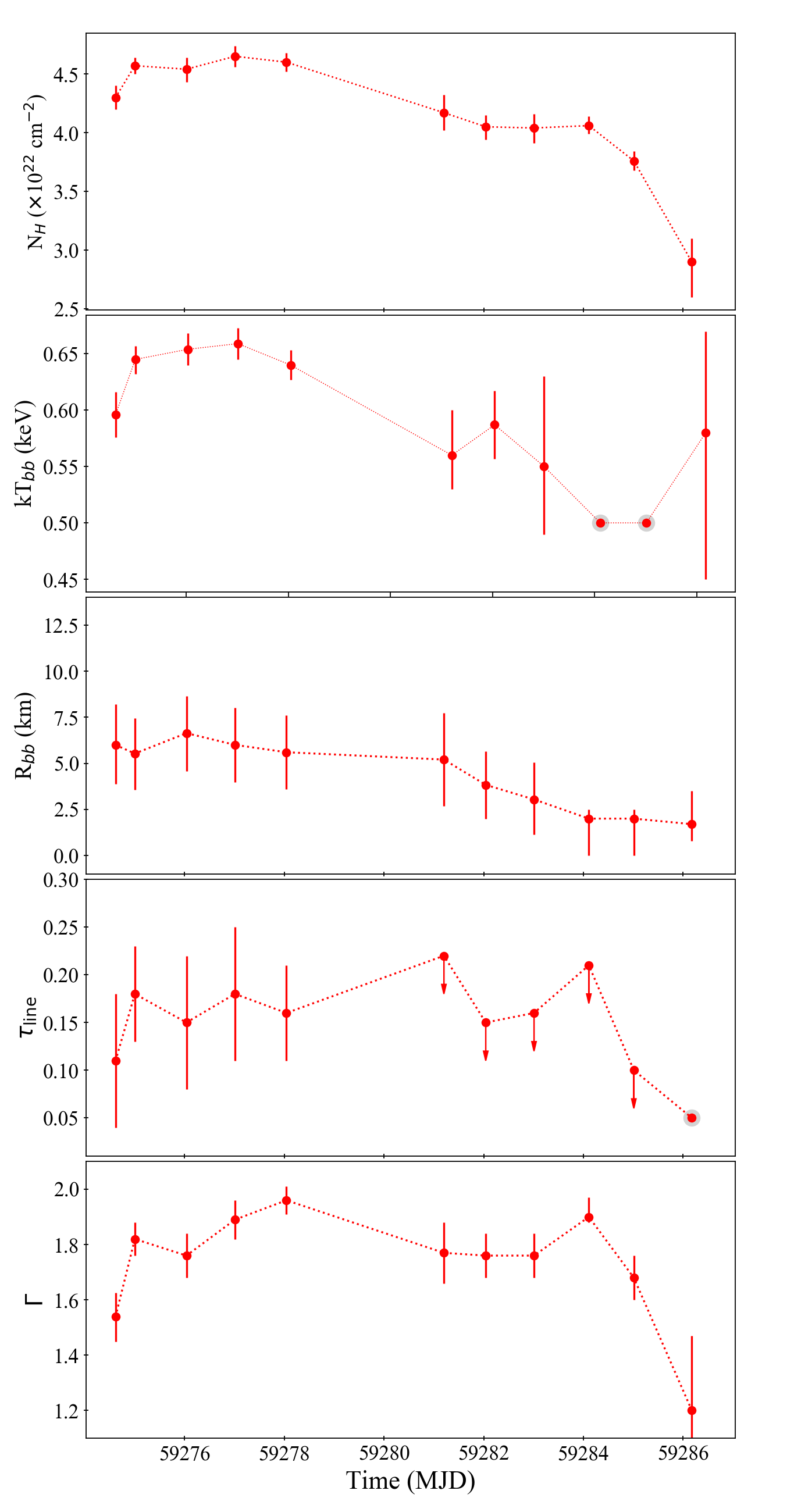}
\caption{Evolution of the best-fit parameters for $N_{\rm H}$, $kT_{\rm bb}$, $R_{\rm bb}$, $\tau_{\rm line}$ and $\Gamma$ over the eleven analysed epochs. Upper (lower) limits are represented with arrows. When corresponding to parameters that were kept frozen, the data points are represented with a gray exterior  outer circle. }
\label{fig:parameters}
\end{figure}

In order to follow the evolution of the single physical parameters along the two-weeks outburst, we analysed each \nicer\ spectrum separately. In the following, we will refer to each of them with labels going from N01 to N11 (see Table \ref{tab:obs}), in progressive order. We modelled the data in accordance with the results obtained in Sec. \ref{ss:broadband} with the same continuum model, i.e., \textsc{tbabs}$\times$\textsc{zxipcf}$\times$(\textsc{bbodyrad}+\textsc{relxillCp}). However, due to the \nicer\ energy range, we kept fixed $kT_{\rm e}$ to 12~keV, in consistency with the results of the broadband spectral analysis. Furthermore, the reduced energy range with respect to the broadband spectrum forced us to fix more parameters, as they were left completely unconstrained in the individual fits. This is the case, in particular, for $f_{\rm refl}$, $R_{\rm in}$,  $N_{\rm H, IA}$, $f_{\rm IA}$ and $\log{\xi_{\rm IA}}$, kept frozen to the (previously obtained) values of 0.1, 30 $R_{\rm G}$, 500 cm$^{-2}$, 0.10 and 4.0 respectively. 
%$f_{\rm IA}$ and $\xi_{\rm IA}$, the main parameter of the model is the equivalent hydrogen column of the material, $N_{\rm H, IA}$.
The presence of discrete features around 1.7~keV in emission and around 7~keV in absorption could be easily spotted in the residuals of observations N01--N06, i.e., around the peak of the outburst. In analogy with Section \ref{ss:broadband}, they were fitted with a $\textsc{gaussian}$ and \textsc{gabs}, respectively. However, for the absorption feature, the estimated errors on $E_{\rm line}$ and ${\sigma}_{\rm line}$ were too large to be meaningful in all spectra, with the exception of observation N04, i.e. the same observation used for the broadband spectral analysis. We therefore fixed those parameters to 7.0 and 0.10~keV, in analogy with the broadband spectrum. The best-fit parameters are reported on Tab. \ref{tab:fit_nicer_free}. Despite the large errors, the optical depth $\tau_{\rm line}$ associated with the feature shows an evolution, i.e. going from 0.15--0.20 in the observations at the outburst peak (N01--N05) to likely smaller values in the fading part of the outburst, where only upper limits could be posed.
%an evolution can be discerned in the optical depth $\tau_{\rm line}$ associated with the feature, which results higher around the peak, i.e., it is comprised between 0.9 and 1.4 for obs. N02--N05, while it ranges between 0.2 and 0.7 for obs. N01 and N06.  %\simo{Forse mi sono persa qualcosa, ma non è molto chiaro a cosa si riferisca il termine "epoch", forse conviene "definirlo" da qualche parte?} \\
In the first six observations, N$_H$ attained values in the range 4.4--4.6$\times$10$^{22}$ cm$^{-2}$, while starting with N07, a decreasing trend could be observed in the best-fit values obtained for N$_{\rm H}$, i.e., going down to $\sim$3$\times$10$^{22}$ cm$^{-2}$ in the last observation.  Fixing N$_{\rm H}$ to the previously obtained value of $\sim$4.5$\times$10$^{22}$ cm$^{-2}$ provided acceptable fits, but with slightly worse $\chi^2$/d.o.f. ratios. In order to check whether such an evolution was significant, we followed the same procedure illustrated in Section \ref{ss:broadband} to calculate the marginal posterior distributions of the best-fit parameters. The results for four representative observations are presented in Figure \ref{fig:corner}. As apparent from the Figure, the changes in N$_H$ between different observations are significant, highlighting that N$_H$ should be indeed left free. In the following, we will therefore report only the results obtained for N$_H$ thawed, and discuss the physical implications of that scenario in Sec. \ref{sec:disc}. The evolution of the main physical parameters of the system is displayed in Fig. \ref{fig:parameters}.
%We then improved the precision of these results ap- plying a Goodman-Weare algorithm of Monte Carlo Markov Chain (MCMC; Goodman & Weare 2010), appropriate in the case of a skewed and elongated distribution for mass and radius, using the command CHAIN in XSPEC. Whenever the mass value was frozen, the MCMC procedure was launched with 300 walkers and a chain length of 9000

%We present the joint and marginal posterior distributions of the accretion disc parameters obtained in the SED fit in Fig. B1 (see Sec. 3.2). We used corner.py (Foreman- Mackey 2016) to visualise the MCMC chains.

%However, the improvements in the $\chi^2$ brought by leaving N$_H$ free are not statistically significant, i.e., of the order 1--1.5$\sigma$, with $\sigma$ considered as $\Delta \chi^2$/$\sqrt{2\nu}$ and $\nu$ the number of degrees of freedom. 

 %, where the data points corresponding to the fits where N$_H$ was kept frozen (free) are identified with squares (circles) for clarity. 
At the beginning of the outburst, the thermal emission is due to hot spots of about 5--6~km in size and $\sim$0.6~keV temperature. Towards the peak of the outburst (around observation 5), $\Gamma$ is characterised by an increasing trend, despite never becoming softer than $\Gamma\sim$2.1. After the peak, at a 0.5--10 keV X-ray luminosity of only 2$\times$10$^{36}$\ergs, the system starts fading to quiescence. During this decay, $kT_{\rm bb}$ decreases slightly (around 0.5~keV) and the hot spot shrinks to about 1--2~km radius. Concerning $\Gamma$, a hardening in the power-law index can be observed while the system fades towards quiescence.

%the paths followed by $kT_{\rm bb}$ and $R_{\rm bb}$ change dramatically depending whether N$_H$ was left free or frozen. In the first case, $kT_{\rm bb}$ decreases slightly (around 0.5~keV) and the hot spot shrinks to about 1-2~km radius. Alternatively, for N$_{\rm H}$ frozen, a much more pronounced breakdown in $kT_{\rm bb}$ and an increasing trend in $R_{\rm bb}$ can be observed. Concerning $\Gamma$, both scenarios imply a hardening, which by the way occurs more rapidly for $N_{H}$ left free.
\begin{table*}
\centering
\begin{tabular}{ l l l l l l l l}
\hline 
%\hline
%\multicolumn{5}{l}{\textbf{Spectral analysis}} \\
\hline
  \multicolumn{7}{c}{\large \nicer spectral analysis} \T \B \\
\hline
 \multicolumn{7}{c}{{\bf model:} \texttt{tbabs}$\times$\texttt{zxipcf}$\times$\texttt{gabs}$\times$(\texttt{relxillCp}+\texttt{bbodyrad})} \\
\hline
{Parameters} & N01 & N02 & N03 & N04 & N05 & N06 \T \B \\
\cmidrule(lr){2-7}
 %\multicolumn{6}{c}{$N_{\rm H}$ free} \\
%\hline
$N_{\rm H}$ \tiny{($\times$10$^{22}$ cm$^{-2}$)}  &  $4.30\pm$0.10 & $4.57\pm$0.07 & $4.54^{+0.11}_{-0.10}$ & {4.65$\pm$0.09} &  $4.60\pm$0.08 & 4.17$\pm$0.15 \T \B \\
$kT_{\rm bb}$ (keV) &  $0.596\pm{0.020}$ & $0.645^{+0.013}_{-0.012}$ & $0.654\pm0.014$ & $0.659^{+0.013}_{-0.012}$ & $0.640\pm{0.013}$ & 0.560$^{+0.030}_{-0.040}$ \\
$R_{\rm bb}$ (km) &  6.0$\pm$2.0 & 5.5$\pm$2.0 & 6.6$\pm$2.0 & 6.0$\pm$2.0 & 5.6$\pm$2.0 & 5.2$\pm$2.5 \\
$\Gamma$ &  1.54$^{+0.09}_{-0.08}$ & 1.82$\pm{0.06}$ & 1.76$\pm{0.08}$ & 1.89$\pm$0.07 & 1.96$\pm$0.05 & $1.77\pm 0.11$ \\
$\tau_{\rm line}$ &  0.11$\pm$0.07 & 0.18$\pm$0.05 & 0.15$\pm$0.07 & 0.18$\pm$0.07 & 0.16$\pm$0.05 & <0.21  \\
$\chi^2_\nu$ (d.o.f.) &  1.08(106) & 0.92(118) & 1.03(109) & 1.08(115) & 1.06(118) & 1.01(100) \T \B \\
$F_{X}^{a}$ \tiny{($\times 10^{-10}$ erg cm$^{-2}$ s$^{-1}$} & 1.944$^{+0.050}_{-0.014}$ & 2.805$\pm$0.012 & 3.430$^{+0.020}_{-0.019}$ & 3.115$\pm$0.013 & 3.112$\pm$0.012 & 1.843$^{+0.019}_{-0.018}$ \\
\cmidrule(lr){2-7}
\\
%\hline
 & N07 & N08 & N09 & N10 & N11 & \T \B \\
\cmidrule(lr){2-6}
%\hline
$N_{\rm H}$ \tiny{($\times$10$^{22}$ cm$^{-2}$)}  &  $4.05^{+0.11}_{-0.10}$ & $4.04^{+0.13}_{-0.12}$ & $4.06^{+0.07}_{-0.08}$ & $3.76\pm 0.08$ & $2.90^{+0.30}_{-0.20}$ \T \B \\
$kT_{\rm bb}$ (keV) & 0.587$\pm$0.030 &  $0.550^{+0.080}_{-0.060}$ & (0.500) & (0.500) & 0.580$^{+0.090}_{-0.130}$ \\
$R_{\rm bb}$ (km) & 3.8$\pm$1.8 & 3.0$^{+1.9}_{-2.0}$ & <2.5 & <2.5 & 1.7$^{+1.8}_{-0.9}$\\
$\Gamma$ & $1.76\pm 0.08$ & 1.76$\pm$0.08 & $1.90^{+0.02}_{-0.07}$ & $1.68\pm 0.08$ & < 1.47 \\
$\tau_{\rm line}$ & <0.15 & <0.16 & <0.21 & <0.11 & (0.05)  \\
$\chi^2_\nu$ (d.o.f.) &  1.00(109) & 1.12(106) & 0.98(98) & 1.08(94) & 1.05(63) \T \B \\
%\hline
%\\
%\hline
 $F_{X}^{a}$ \tiny{($\times 10^{-10}$ erg cm$^{-2}$ s$^{-1}$} & 1.594$\pm$0.010 & 1.277$\pm$0.009 & 0.840$^{+0.009}_{-0.008}$ & 0.691$\pm$0.009 & 0.300$^{+0.008}_{-0.007}$ \\
%
%\tiny{($\times$10$^{22}$ cm$^{-2}$)} 
%Obs. & $N_{\rm H}$ &  $kT_{\rm bb}$ & $R_{\rm bb}$ & $\Gamma$ & $\tau_{\rm line}$ & $F_{\rm 0.5-10 \ keV}$ & $\chi^2_\nu$ \T \B \\
%\hline
%&  ($\times$10$^{22}$ cm$^{-2}$)  & (keV) & (km) &  & & \tiny{($\times 10^{-10}$ erg cm$^{-2}$ s$^{-1} $)} & (d.o.f.) \T \B \\
\cmidrule(lr){2-6}

\\
\hline
\hline
\end{tabular}
%\end{adjustbox}
\caption{Results of the spectral analysis of the single \nicer spectra. Quoted errors reflect 90 \% confidence level. The parameters that were kept frozen during the fits are reported between round parentheses. $^a$: The flux values reported, corresponding to the 0.5--10~keV energy range, have been obtained in the N$_H$ free case.}
\label{tab:fit_nicer_free}
\end{table*}

%Some parameters were frozen, as the lack of data beyond 10~keV made the fit unable to find any constraint on them. We did not include all the lines detected in the broadband spectrum, but rather added components if specifically required by residuals inspection. Spectral shape was found quite stable between the subsequent data sets (see fig. \ref{fig:parameters}). 
%\paragraph{N$_{\rm H}$ free} 

%The values of \kt attained a value of about 0.5~keV, with a cooling trend from observation *05. $\Gamma$ was rather constant as well.

%\subsection{Cumulative \nicer spectrum}\label{ss:nicer_sum}

\section{Discussion}\label{sec:disc}
In this manuscript we presented the spectral analysis of \source\ during its 2021 faint, two-weeks lasting outburst. We used data from the \nicer daily monitoring of this outburst and included \xmm and \nustar data taken at the peak. As in several other AMXPs \citep[see table 2 of ][for an overview of the outburst properties of the class]{Marino2019a}, the system displayed a relatively short and hard outburst, with very little spectral evolution. In the following we discuss the main findings of this work and their physical implications. 
\subsection{Geometry of the accretion flow}\label{ss:geometry}
Both broadband and individual \nicer spectra were satisfactorily described with a black body component, a thermal Comptonisation spectrum and a reflection component. At the peak, the black body emission arises from a $\sim$5--6~km region at a temperature of about 0.6~keV, consistent with a hot spot on the surface of the NS, possibly the magnetic polar caps. The Comptonisation spectrum is instead due to "cold" seed photons (see Section \ref{ss:broadband}), according to the assumptions underlying the \textsc{relxillCp} model, Compton up-scattered by a 12~keV electron corona into a $\Gamma \sim$1.9 cut-off power-law. The source of these seed photons is probably attributable to the accretion disc, which is indeed too faint to give a statistically significant contribution to the spectrum or simply too cold for the energy range considered. A small, i.e., $f_{\rm refl}\sim$10-20\%, fraction of these photons are intercepted by the disc and reprocessed, forming the reflection component. Modelling this one with \textsc{relxillCp}, we obtained that the disc is truncated quite far from the NS, at a $\sim$20--34~R$_{\rm G}$ (44--76~km for a 1.5 M$_{\odot}$ NS) radius, and it is quite highly ionised, indeed $\log \xi$ is consistent with 3.3. Considering that these estimates should be taken with some caution due to the uncertainties on e.g. the value adopted for the NS mass, the obtained range of values for the inner edge of the disc results roughly compatible with the co-rotation radius $R_{\rm co}$ expected for a NS spinning at 518 Hz, i.e., 31--44~km for a 0.8--2.2 M$_{\odot}$ NS. It results thereby that the inner edge of the disc and the NS magnetic field lines spin at (almost) the same frequencies and the torque exerted on the NS by the columns of matter funneled up onto the magnetic field lines should be negligible. According to this result, the NS in \source\ is expected to be in a spin-equilibrium condition, i.e., not subject to any significant spin-up or spin-down rate. It is noteworthy that the spin derivative obtained in \cite{Sanna2022} are compatible with this scenario. Similar moderate truncation radii have been found in AMXPs that have shown remarkably bright outbursts, such as HETE~J1900.1-2455 \citep[][]{Papitto2013_hete} and SAX J1748.9-2021 \citep[][]{Pintore2016}, but also for the very faint AMXP IGR~J17062–6143 \citep[][]{Bult2021}.
%, for which an inner disc higher than 100~R$_{\rm G}$ has been obtained \citep[][]{Degenaar2016, VanDenEijnden2018}. 
If confirmed, this finding would suggest that high truncation radii are not an essential ingredient to obtain sub-luminous outbursts. \\
%Accretion discs are typically truncated in AMXPs, but the values reported in literature for their inner edges range from $\sim$10 to $\sim$30 R$_{\rm G}$ \citep[see, e.g., ][]{Miller2011,Papitto2013,King2016,Pintore2016,DiSalvo2019}, thereby significantly lower than our estimate. 
%A similar high value has been obtained for another AMXP accreting at very low luminosities, i.e., IGR~J17062–6143, with an inner disc higher than 100 R$_{\rm G}$ \citep[][]{Degenaar2016, VanDenEijnden2018}. \\
%High inner radii of the accretion discs have been invoked to explain the dim luminosity range of VFXTs \citep[e.g., ][]{Heinke2015}. 
If disc truncation is a consequence of the NS dipolar magnetic field halting the accretion flow, the inner edge of the disc obtained from reflection must coincide with such magnetospheric radius, i.e., the distance where the magnetospheric pressure $R_m$ \citep[][]{Ghosh1979}, equals the ram pressure of the accreted matter. 
In order to test this scenario, we solved the following formula \citep[][]{Frank2002}:
\begin{equation}
    R_m = \phi \times 2.9\times10^8 L^{-2/7}_{37} M^{1/7}_{1.4} R^{-2/7}_6 \mu^{4/7}_{30} \ {\rm cm} \ \ ,
\end{equation}
where $\phi$ is a factor of order 0.3--0.5 \citep[][]{Burderi1998} which accounts for disk-fed accretion flows, $L_{37}$ is the bolometric luminosity in 10$^{37}$~\ergs units, $M_{1.4}$ the NS mass in 1.4~M$_\odot$ units, $R_6$ the NS radius in 10$^6$~cm units and $\mu_{30}$ (equal to $BR^3$) is the magnetic dipole momentum in units of 10$^{30}$~G~cm$^{3}$. By assuming $M_{1.5}=1, \ R_6=1$ and $R_m = R_{\rm in}$, we solved for $B$ and found that to justify the truncation radius obtained from reflection, the magnetic field must be comprised between 0.9$\times$10$^{9}$~G and 3.0$\times$10$^{9}$~G, corresponding to $\phi$=0.5. The obtained result is in line with the upper limits obtained for several AMXPs \citep[e.g. ][]{Mucherjee2015} but almost an order of magnitude higher with respect to the upper limit on the magnetic field measured from the long-term NS spin evolution in the same source, i.e. $B<1.3\times10^{8}$ G \citep[][]{Sanna2022}. The discrepancy may indicate that the disc is not truncated by the magnetospheric pressure and that other mechanisms are responsible for halting the disc, e.g., the accretion flow becomes radiatively inefficient beyond such radius as typically observed in black hole X-ray binaries at low luminosities. It is also noteworthy that the reflecting region may not necessarily coincide with the real inner edge of the accretion disk, for example if some self-shielding effects are at play, so that the latter may not coincide with the measure of $R_{\rm in}$ obtained in this analysis. 

\subsection{On the identification of the spectral state}\label{ss:spectral-state}
\begin{figure}
\centering
\includegraphics[width=\columnwidth]{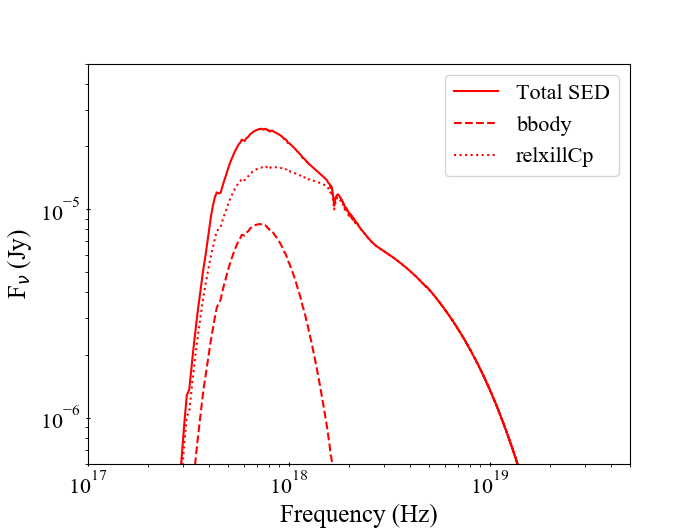}
\caption{\source spectral energy distribution in Epoch 4, determined with the best-fit parameters reported in Table \ref{tab:fit_broadband}. The single spectral components are highlighted with different line styles.}
\label{fig:SED}
\end{figure}

When observed in an outburst, AMXPs are typically hard and rarely display a transition to the soft state \citep[][]{DiSalvo2020}. This behaviour is not found in most (persistent or transient) atolls \citep[][]{Hasinger1989}, which show a variety of states, from hard to soft, characterised by different spectral and timing properties. In \source, the hardness ratio remains quite high, i.e. about 3-5, for the whole outburst, suggesting a behaviour similar to the other AMXPs. Furthermore, the spectral continuum is clearly never dominated by the disc and/or the NS thermal components, as expected instead for soft states. The electron temperature of the corona found in the Epoch 4, the only one for which we have broadband spectral coverage, i.e. 12--14 keV, is also comparable with the temperature found for other NS LMXBs in hard-intermediate states \citep[e.g.,][]{Pintore2018,Marino2019b}. It is noteworthy that the the Spectral Energy Distribution (SED) for Epoch 4 (Fig. \ref{fig:SED}), appears similar to the SED found in other NS LMXBs in hard state \citep[e.g. ][]{Bianchi2017,Ponti2019}, in particular considering the spectral cut-off energy, i.e. beyond 10$^{19}$ Hz. Such a comparison is quite intriguing, since those authors found that winds were absent in the observations characterised by such hard SEDs but were present instead in softer states. However, the timing properties of \source would be more reminiscent of a soft state and somehow contradict the hard state identification. NS LMXBs in hard state display typical values of {\it rms} amplitude of about 10-20 \% or higher \citep[see, e.g. ][]{MunozDarias2014}, while \source seems to show very little variability ({\it rms} of about 6\%). It is noteworthy that such a lack of variability characterizes all \nicer observations, also the ones taken at the end of the outburst, where the hardness ratio increase and the $\Gamma$ decreasing trends would suggest a hardening. This points out that the timing properties of the system are rather anomalous. However, a comprehensive study on the origin of such weak variability goes beyond the scope of the paper. Finally, we note that the difficulty in detecting any power in the PDS could be partly due to \nicer being most sensitive at low energies, where low X-ray variability is expected.

\subsection{Detection of disc winds out of a canonical soft state}\label{ss:wind}
The detection of an absorption blueshifted Fe XXVI line at $\sim$7.0~keV indicates the presence of a disc wind in the outburst. According to the estimated outflow velocity, matter is ejected at 600--2700 km/s ($\sim$0.002--0.007~c). The feature was found also in six out of eleven single \nicer observations, although statistics in those spectra is too low to find any constraints on the centroid energy and its width. No trace of similar structures could be spotted in the residuals of the remaining five spectra and we were able only to estimate upper limits on the absorption strength and on the equivalent width (upper limit of about 0.02 keV or lower in all observations). In order to establish whether the disappearance of the absorption line was intrinsic or due to the lower statistics, we summed these five spectra together, finding that the feature was absent also in such joint spectrum. Only upper limits could be estimated for the absorption strength of such feature, corresponding to $\tau_{\rm line}$<0.04, and to the equivalent width, lower than 0.02 keV also in this case. Such test points out that the outflow could be active only during the peak of the outburst. Indeed, the further hardening of the continuum in the final stages of the outburst, witnessed e.g. by a generally decreasing $\Gamma$ in Table \ref{tab:fit_nicer_free}, could make the wind unstable and be responsible for its disappearance \citep[see, e.g. ][]{Chakravorty2013, Higginbottom2020, Petrucci2021}. \\
Disc winds are commonly observed in X-ray binaries viewed at high inclination harbouring both BHs \citep[see, e.g., ][]{Ponti2012} and NSs \citep[see, e.g., ][]{Dai2014, Pintore2014}, with orbital periods typically large, i.e., $\sim$~hrs \citep[][]{DiazTrigo2016}. According to the standard observational picture in BH systems \citep[see, e.g., ][]{Miller2008}, winds are typically observed only during soft states, while matter ejection in hard state occurs mainly in the form of compact jets \citep[but see, ][for a critical discussion on jet-winds mutual exclusion]{Homan2016}. The detection of winds only in the soft states of the NS LMXBs EXO 0748--676 \citep[][]{Ponti2014} and AX J1745.6--2901  \citep[][]{Ponti2015} suggest a similar pattern for NS systems. However, in the last few years, the detection of optical/UV winds in several transient systems during their hard state, e.g., in the BH LMXBs MAXI~J1820+070 \citep[][]{MunozDarias2019}, MAXI~J1803-298 \citep[][]{MataSanchez2022} and recently in the NS XRB Swift J1858.6--0814 \citep[][]{CastroSegura2022}, has complicated the picture, revealing that outflow signatures in the optical/UV band may be common features in XRBs in hard states. AMXPs pose further challenges to our understanding of when and where disc winds are expected, as several claims for Doppler-shifted absorption features in the X-ray band have been put forward. The discovery of a Doppler-shifted Si XIII line at $\sim$1.9 keV has been firmly detected in IGR~J17591--2342 \citep[][]{Nowak2019} and interpreted as the signature of disc winds at $\sim$0.1~$c$. Weak detections of outflows have been reported also in two other AMXPs \citep[][]{VanDenEijnden2018, DiSalvo2019}, which, if confirmed, would also imply the existence of winds in systems with orbital periods lower than 2 hr. The discovery of a weakly relativistic equatorial wind in \source\ may represent another clear exception to the aforementioned winds paradigm. Winds are indeed present in an observation which, based on the spectral characteristics, would be more consistent with an intermediate or hard-intermediate state (see Sec. \ref{ss:spectral-state}). Concerning the origin of such outflow, among the three physical mechanisms typically invoked for winds, i.e., thermally-driven, magnetically-driven and radiation pressure-driven \citep[see, e.g., ][and references therein]{DiazTrigo2016}, the latter can not hold in the present case as it would require Eddington or super-Eddington luminosities \citep[][]{Proga2002}, unless the intrinsic luminosity in the system is higher and scattering from the surrounding material makes it look fainter \citep[as in the case of 4U 1822--37, see e.g.][]{Anitra2021}. In order to distinguish between thermally and magnetically-driven winds, the radius of wind launching should be constrained, since thermal winds are expected to arise from the outermost part of large discs, i.e., where the sound speed of the plasma overcomes the Keplerian speed \citep[see, e.g., ][]{Higginbottom2017}. It is noteworthy that in magnetised NSs, winds could also be in principle propeller-driven, i.e., launched when the magnetic field is rotating faster than the accretion flow creates a centrifugal barrier able to sweep matter away \citep[e.g., ][]{Illarionov1975, Romanova2009}. As already discussed in Section \ref{ss:geometry}, the high truncation radius obtained from reflection could be ascribed to the magnetospheric pressure dominating over ram pressure at such distance and this, in turn, could be responsible for wind launching in this particular case. The onset of propellers require a dipole magnetic field interacting with the accretion flow and a relatively weak accretion rate, both ingredients typically present in AMXPs, but absent in BH LMXBs and in the other, non-pulsating NS LMXBs (where the magnetic field could be buried inside the NS) as well, and could constitute an explanation for observing disc winds even outside of the spectral soft state.  \\
Along with the evidence of a wind, we also included a model describing absorption by ionised material, i.e., \textsc{zxipcf}, but only in the broadband spectrum the parameters of the model could be constrained. According to the obtained best-fit parameters, such material covers only $\sim$10\% of the background illuminating X-ray source and it has high column density ($N_{H, IA} \gtrsim 2.9\times10^{24}$ cm$^{-2}$) and high ionization state ($\log{\xi_{IA}}\sim3.1-4.4$). According to the low covering fraction found, this absorbing material could be interpreted as an ionised layer of plasma coating the optically thick disc and seen edge-on, most likely associated with the Fe XXVI wind. It is noteworthy that with such a high column density, absorption features produced by lower Z species should appear at soft X-rays, but this effect is masked by the large neutral column density $N_{\rm H}$ due to interstellar absorption. Furthermore, the highly dense local ionised absorber could also produce significant Thomson scattering, which in turn would reduce the X-ray luminosity and in principle explain the faintness of the system.
Unfortunately, the parameters of the model in the single \nicer spectra were too unconstrained to check for an evolution of such absorbing material along with the outburst decay.

%Winds have been firmly detected in another AMXP \citep[][]{Nowak2019}, and suggested in other two sources \citep[][]{VanDenEijnden2018, DiSalvo2019}, all of them in hard states. It is also noteworthy that winds have been recen
%Interestingly they are usually found in soft state. This is quite a hard state. %Timing results by Sanna et al., 2021 in prep.

\subsection{Outflows and non-conservative mass-transfer}
Matter eruption in the form of winds is able in principle to remove a significant amount of mass from the system. The mass-loss rate due to winds, $\dot{M}_w$, can be indeed expressed as \citep[][]{Ponti2012,Ponti2015}:
\begin{equation}\label{eq:mdot_wind}
    \dot{M}_w=4\pi m_p v_{\rm out} \frac{L_X}{\xi} \frac{\Omega}{4\pi} \ \ \ \ ,
\end{equation}
with $m_p$ the proton mass, $v_{\rm out}$ and $\xi$ the velocity and the ionization of the outflowing plasma, respectively, and $\Omega$ the solid angle subtended by the wind. Since we do not detect spectral signatures of Fe XXV or other less-ionised species of iron, the plasma is most likely highly ionised and we assume $\xi$ to be 10$^4$, in accordance with the ion distributions computed by \cite{Kallman2001} and the values obtained for the ionisation of the absorber with \textsc{zxipcf}. 
In order to calculate $\Omega$, we considered two cases: we first assumed a wind opening angle of 30$^\circ$ \citep[as in e.g.][]{Ponti2012} or 10$^\circ$, the latter under the hypothesis that the wind and the ionised material (covering about 10\% of the X-ray source, see Sec. \ref{ss:broadband}), are associated with each other. Finally, at the outburst peak, i.e. from the analysis of the broadband spectrum, we estimated $v_{out}$ in the range 600-2700 km s$^{-1}$ and a bolometric X-ray luminosity of $\sim$7$\times$10$^{36}$ erg s$^{-1}$, which corresponds to a $\dot{M}_w$ of 0.6--2.7$\times$10$^{17}$ g s$^{-1}$ for $\Omega=30^\circ$ and 0.7--3$\times$10$^{16}$ g s$^{-1}$ for $\Omega=10^\circ$. Now, considering that the mass-accretion rate $\dot{M}_2$ and $L_X$ are connected by the relation $L_X=\frac{GM_1\dot{M}_2}{R_1}$, with $M_1$ and $R$ NS mass and radius and assuming a NS mass of 1.5 M$_\odot$ and radius of 12 km \citep[values commonly found for NSs, see ][for a review]{Ozel2016}, we estimated the rate of mass inflow at the outburst peak to be about $\sim$4$\times$10$^{17}$ g s$^{-1}$. It follows that, for $\Omega$ $\dot{M}_w \sim 0.15-0.70 \dot{M}_2$, i.e. the rate of mass outflows due to winds is almost comparable to the rate of mass inflow. The overall mass-transfer regime results to be therefore definitely non-conservative. \\
A highly non-conservative mass-transfer (NCMT, hereafter) regime\footnote{But see also \cite{Applegate1994} for an alternative theory.} has been invoked in the past to explain the strong orbital expansion rates, too strong to be explained with angular momentum losses via gravitational radiation or magnetic braking, observed in (almost) all the AMXPs observed more than once in outburst \citep{Disalvo2008, Burderi2009a, Mucherjee2015, Sanna2018_1737, Bult2021}. According to the orbital period derivative of \source , measured after the outburst analysed here, the system is rapidly expanding as well, and also in this case such phenomenon could imply a NCMT \citep[][]{Sanna2022}. A high rate of mass-loss was also implied by multi-wavelength modelling of IGR~J17062--6143 \citep[][]{HernandezSantisteban2019}, somehow giving indirect evidence for the outflow claimed by \cite{VanDenEijnden2018}. The (unexpected) presence of winds (see Sec. \ref{ss:wind} and references therein) and the evidence that AMXPs appear more radio bright than the other non-pulsating NS LMXBs, i.e., they exhibit stronger ejections in the form of jets \citep[e.g., ][]{Tetarenko2016, Russell2018}, corroborate a scenario where AMXPs are very efficient engines when it comes to ejection of significant amounts of the mass transferred by the companion. \\ 
Indirect evidences for a NCMT scenario in almost half of the AMXPs known so far have been found from the comparison between the energy output expected in the conservative case, i.e. where all the transferred mass is effectively accreted onto the NS surface, and the energy output actually observed. Such a methodology was developed in \cite{Marino2017} and then applied to the large majority of the known AMXPs in \cite{Marino2019a}. These authors obtained strong indications for a NCMT in five systems in the sample, including \source . In this paragraph, we apply this method again to our system, including three additional years of quiescence and the 2021 outburst, analysed in this manuscript. Broadly speaking \citep[see][for a detailed description of the method and discussion of the main caveats]{Marino2019a}, the method consists of comparing the total amount of energy emitted by the system $E_{\rm tot,exp}$ expected in the case of a fully conservative mass-transfer with the observed energy output $E_{\rm tot, obs}$ measured during each outburst displayed by the system since 1996\footnote{Since 1996, the X-ray sky was almost continuously monitored, so that we had good chances of recording any outburst from an AMXP.}. Considering that the expected luminosity $L_{\rm exp}$ can be expressed as $L_{\rm exp}=\displaystyle\frac{GM_1\dot{M}_2}{R}$, with $M_1$ and $R$ NS mass and radius and $\dot{M}_2$ the mass-transfer rate, $L_{\rm exp}$ was obtained by adopting $M_1=1.5 M_\odot$ and $R_1$=12~km. In order to compute $\dot{M}_2$, we solved equations (2) and (3) in \cite{Marino2019a}, using for $M_2$ the value derived from the mass function $f=0.055$ by fixing the inclination to the known value of 77$^\circ$ \citep{Altamirano2011_1749} and considering only angular momentum losses via gravitational radiation (GR). The total expected energy output $E_{\rm tot, exp}$ can be obtained by simply multiplying $L_{\rm exp}$ by the chosen baseline, i.e. the 25 years period between 1996 and 2021, and it equals $E_{\rm tot, exp}=$1.3$\times$10$^{44}$ ergs. Concerning the observed energy output $E_{\rm tot, obs}$, for both the 2010 outburst \citep[see][for more details]{Marino2019a} and the 2021 outburst analysed here, we estimated the fluence ($\phi_{2010}$ and $\phi_{2021}$) as the area subtended by the X-ray light curve of each outburst and extended this value in the range 0.5--300~keV. By taking 6.7 kpc as the distance of the system, the value of $E_{\rm tot, obs}$ is obtained by scaling the fluence for the area over which such fluence spreads before reaching the observer, i.e. $E_{\rm tot,obs}=(\phi_{2010}+\phi_{2021})\times4\pi d^2$. The observed energy output results therefore equal to $\sim$1.5$\times$10$^{43}$, i.e. one order of magnitude below the expected energy for a conservative scenario. Even considering the contribution of the energy emitted during quiescence, at an average X-ray luminosity of $10^{33}$-$10^{34}$ \ergs , the two values could not be reconciled and, unless we missed one or more outburst during the time taken into account, the previous claim that the system is undergoing a NCMT scenario \citep[][]{Marino2019a} is confirmed. We also notice that the companion star in \source is not a degenerate helium dwarf or a white dwarf, so that the effects of magnetic braking should be included. Considering this additional angular momentum loss channel would make the expected luminosity even higher, so that the value obtained for $E_{\rm out, exp}$ has to be considered a lower limit. \\
Whether or not the disc winds found in this work could be entirely responsible for the discrepancy between expected and observed luminosity discussed above and the high orbital period derivative measured in the system \citep[][]{Sanna2022}, it is unclear. In order to investigate such connection, we should know how much angular momentum winds extract from the system, i.e. the location from where winds are launched, and we should have estimates of $\dot{M}_w$ for the whole outburst, not only at the peak. It is also most likely that the NCMT observed results from several channels of mass-loss other than winds, such as the removal of matter during quiescence due to pulsar winds \citep[see, e.g.][]{Burderi2001,Parfrey2017} and jets.

\subsection{The end of the outburst: dawn of a power-law quiescence spectrum?}\label{ss:disc-thermal}
In Section \ref{ss:nicer_monitoring}, we showed that, past the outburst peak, N$_H$ decreases significantly, the hot spot shrinks, going from 6 to 2~km radii, and the continuum hardens.
%In Section \ref{ss:nicer_monitoring}, we found that the choice of fixing or rather leaving free N$_H$ generates two possible evolutionary paths for the temperature and size of the black body emitting region during the outburst decay. For N$_H$ constant, the hot spot increases in size to such an extent that it may become not compatible with the whole NS surface anymore. On the contrary, allowing N$_H$ free, the hot spot shrinks, going from 6 to 2~km radii. The evolution of these black body components in the two scenarios is displayed in Fig. \ref{fig:bbody}. In both cases, the spectrum hardens, although more rapidly in the variable-N$_H$ scenario. \\
A scenario with variable N$_H$ can be explained by invoking changes over short time-scales of local, neutral absorption in the system. Concerning the decreasing trend in the black body radius, several mechanisms may be invoked to explain it. It is plausible that the decrease in mass-accretion rate throughout the outburst implies a progressively smaller amount of energy supplied to the polar caps through the magnetic field lines. Along with the energy, we suggest that the area over which this energy is distributed may decrease as well. A similar reduction in the hot spot size at the end of the outburst could be a consequence of the disc being truncated further away, e.g. due to the reduced mass-accretion rate. Indeed, as the material moves outwards, it pulls away the coupled magnetic field lines, forcing their footprints on the neutron star surface to move inwards and thereby reducing the hot spot perimeter. However, whether such an evolution is compatible with the increase of the pulsed fraction in the final stages of the outburst is yet to be established \citep[][]{Sanna2022}. \\ Unfortunately, the lack of data at soft X-rays taken during the (relatively unexplored) termination of X-ray binary outbursts makes drawing comparisons with analogous similar sources quite challenging. However, it is noteworthy that in the "very faint", i.e., 10$^{34}$-10$^{36}$~\ergs , and quiescence, below 10$^{34}$~\ergs , luminosity ranges, AMXPs tend to show somehow harder spectra \citep[with some exceptions, e.g.][]{Ng2021}, i.e., typically dominated by power-law like spectral components, with respect to other NS LMXBs \citep[see, e.g.][and references therein]{Campana2005, Degenaar2012, Linares2014, Wijnands2015}. The origin of this power-law shaped spectrum in quiescence has been ascribed to residual accretion or physical processes related to the NS magnetic field \citep[e.g., ][]{Fridriksson2011, Chakrabarty2014, Parikh2017}. This occurs in striking contrast with respect to a large number of NS LMXBs, typically non-magnetised, that show purely soft, quasi-thermal spectra, due to the slow release of the energy stored in the crust when accretion was ongoing \citep[see, e.g., ][]{Brown1998,Degenaar2011,Servillat2012,Marino2018}. When observed in quiescence in 2011, \source X-ray spectrum was found to be well described with a simple power-law model \citep{Degenaar2012}, without any contribution by a thermal component. It is therefore plausible that the hardening of the spectrum and the shrinking of the hot spot at the end of the outburst in \source witness the dawn of a power-law dominated quiescence spectrum. Future observations of the system in quiescence are necessary to confirm this hint.

\section{Conclusions}\label{sec:concl}
In this manuscript, we investigated the X-ray emission of the AMXP \source during its short and rather dim outburst that occurred in 2021. Exploiting data from {\it XMM-Newton}, \nustar and \nicer at the outburst peak and from \nicer in the whole outburst, we were able to characterise the geometry of the accretion flow and the main physical parameters of the system and to scrutinise their evolution from rise to decay. \\
In the following, we summarize the main findings of our work:
\begin{itemize}
    \item Throughout the outburst, the spectrum can be described by the sum of a black body component, likely due to a hot spot on the NS surface, and a Comptonisation spectrum which can be associated with a hot corona scattering off photons from a cold disk, whose direct emission is likely too faint for detection;
    \item From the analysis of the reflection component, we estimated the radius of the accretion disc to be about 20--34 \rg (40--68~km). A magnetic field strength of at least $9\times10^8$ G would be required if this radius coincides with the magnetospheric radius, i.e., the interruption of the disc is caused by the magnetospheric pressure taking over.
    \item A blueshifted Fe XXVI line was found in absorption in the broadband spectrum and in six out of eleven \nicer spectra. The presence of this feature witnesses the presence of a mildly relativistic equatorial disc wind, at a velocity of about 0.2--0.7\% $c$. Despite winds being typically thought to be present in soft states only, we point out that the detection of these outflows in a number of AMXPs, including \source, could suggest a different launching mechanism, e.g., a propeller-driven wind.
    \item An additional spectral component, taking into account absorption from an ionised plasma, is required by the fit. As such plasma covers only 10\% of the source, it is compatible with a thin layer of ionised plasma coating the disc, most likely associated with the detected disc wind.
    \item We were able to follow the evolution of the hot spot on the NS surface until the end of the outburst and observed it cooling down and shrinking down to a size of $\sim 2$~km$^{2}$ as it faded to quiescence. This may be consistent with the quiescence spectrum of the system being a pure power-law spectrum, with negligible contribution from the NS surface.
    \item Following \cite{Marino2019a}, where a method to check whether the mass-transfer in a LMXB is conservative or non-conservative was developed and applied to a sample of 19 AMXPs, we updated the results obtained for \source including the outburst analysed in this paper. Although the discrepancy between expected average luminosity for a conservative mass-transfer and the observed one is lower than in our previous work, it is still quite large and confirms that \source loses a significant fraction of the mass transferred by the companion.
\end{itemize}
Despite many outbursts being surveyed in detail in the {\it RXTE} era, with \nicer we have the unprecedented opportunity to follow how the system evolves in the soft X-ray band, which permits accessing e.g., the details of the black body emission. Furthermore, thanks to \nicer sensitivity and flexibility, we were able to follow the whole outburst until its transition to quiescence, when the system enters in a faint and relatively poorly explored flux regime (i.e., below 10$^{36}$~erg~s$^{-1}$). Further studies of this and/or analogous objects will give the opportunity to investigate the behaviour of accreting NSs at low Eddington rates and the critical role of matter outflows in such states. \\

%What we found in this paper. Interesting point: with NICER we are able for the first time to follow the evolution of an outburst in the soft X-rays band. More surveys of outbursts of pulsating and non-pulsating NS LMXBs are necessary to draw conclusions on the evolution of these objects during an outburst.

\section*{Acknowledgements}
We would like to thank the anonymous referee for their useful comments. AM acknowledges a financial contribution from the agreement ASI-INAF n.2017-14-H.0 and from the INAF mainstream grant (PI: T. Belloni, A. De Rosa). AM, TDS, AA and RI acknowledge financial contribution from the HERMES project financed by the Italian Space Agency (ASI) Agreement n. 2016/13 U.O. AM is supported by the H2020 ERC Consolidator Grant “MAGNESIA” under grant agreement No. 817661 (PI: Rea) and National Spanish grant PGC2018-095512-BI00. This work was also partially supported by the program Unidad de Excelencia Maria de Maeztu CEX2020-001058-M, and by the PHAROS COST Action (No. CA16214). PB acknowledges support from the CRESST II cooperative agreement (80GSFC21M0002). SG acknowledges the support of the Centre National d'Etudes Spatiales (CNES). GCM was partially supported by Proyecto de Investigación Plurianual (PIP) 0102 (Consejo Nacional de Investigaciones Científicas y Técnicas (CONICET)) and by PICT-2017-2865 (Agencia Nacional de Promoción Cient\'ifica y Tecnológica (ANPCyT)). DA acknowledges support from the Royal Society.

%The Acknowledgements section is not numbered. Here you can thank helpful
%colleagues, acknowledge funding agencies, telescopes and facilities used etc.

%%%%%%%%%%%%%%%%%%%%%%%%%%%%%%%%%%%%%%%%%%%%%%%%%%
\section*{Data Availability}

The data utilised in this article are publicly available at \url{https://heasarc.gsfc.nasa.gov/cgi-bin/W3Browse/w3browse.pl}, while the analysis products will be shared on reasonable request to the corresponding author.

%The inclusion of a Data Availability Statement is a requirement for articles published in MNRAS. Data Availability Statements provide a standardised format for readers to understand the availability of data underlying the research results described in the article. The statement may refer to original data generated in the course of the study or to third-party data analysed in the article. The statement should describe and provide means of access, where possible, by linking to the data or providing the required accession numbers for the relevant databases or DOIs.



\bibliographystyle{mnras}
\bibliography{biblio} % if your bibtex file is called example.bib
%\appendix

%\section{Spectral fitting of the single \nicer observations}
%In this Appendix, we report on the results of the spectral analysis of the single \nicer spectra, obtained by keeping $N_{\rm H}$ free (Table ...) or frozen (Table ...). 

%\begin{thebibliography}{99}
%\end{thebibliography}

%%%%%%%%%%%%%%%%%%%%%%%%%%%%%%%%%%%%%%%%%%%%%%%%%%

%%%%%%%%%%%%%%%%% APPENDICES %%%%%%%%%%%%%%%%%%%%%

%\appendix



%%%%%%%%%%%%%%%%%%%%%%%%%%%%%%%%%%%%%%%%%%%%%%%%%%

% Don't change these lines
\bsp	% typesetting comment
\label{lastpage}
\end{document}